\documentclass[12pt,preprint]{aastex6}

\usepackage{epsfig}

\usepackage[caption = false]{subfig}

\usepackage{amsmath}
\usepackage{graphicx}
\usepackage{float}
\usepackage{amsfonts}
\usepackage{amssymb}
\usepackage{makeidx}
\usepackage{mathrsfs}
\usepackage{flushend}

\newcommand{\sigmaT}{\sigma_{_{\rm T}}}
\newcommand{\lapprox}{\lower.4ex\hbox{$\;\buildrel
<\over{\scriptstyle\sim}\;$}}

\newcommand{\gapprox}{\lower.4ex\hbox{$\;\buildrel
>\over{\scriptstyle\sim}\;$}}

\begin{document}

\title{TIME-DEPENDENT ELECTRON ACCELERATION IN PULSAR WIND TERMINATION SHOCKS: APPLICATION TO THE 2011 APRIL CRAB NEBULA GAMMA-RAY FLARE}

\author{John J. Kroon}

\affil{National Research Council, resident at the Naval Research Laboratory; Washington, DC 20375, USA; john.kroon.ctr@nrl.navy.mil; jkroon@gmu.edu}

\author{Peter A. Becker}

\affil{Department of Physics and Astronomy, George Mason University, Fairfax, VA 22030-4444, USA;pbecker@gmu.edu}

\author{Justin D. Finke}

\affil{Space Science Division, Naval Research Laboratory, Washington, DC
20375, USA;justin.finke@nrl.navy.mil}

\begin{abstract}

The $\gamma$-ray flares from the Crab nebula observed by {\it AGILE} and {\it Fermi}-LAT between 2007-2013 reached GeV photon energies and lasted several days. The strongest emission, observed during the 2011 April ``super-flare,'' exceeded the quiescent level by more than an order of magnitude. These observations challenge the standard models for particle acceleration in pulsar wind nebulae, because the radiating electrons have energies exceeding the classical radiation-reaction limit for synchrotron. Particle-in-cell simulations have suggested that the classical synchrotron limit can be exceeded if the electrons also experience electrostatic acceleration due to shock-driven magnetic reconnection. In this paper, we revisit the problem using an analytic approach based on solving a fully time-dependent electron transport equation describing the electrostatic acceleration, synchrotron losses, and escape experienced by electrons in a magnetically confined plasma ``blob'' as it encounters and passes through the pulsar-wind termination shock. We show that our model can reproduce the $\gamma$-ray spectra observed during the rising and decaying phases of each of the two sub-flare components of the 2011 April super-flare. We integrate the spectrum for photon energies $\ge 100\,$MeV to obtain the light curve for the event, which agrees with the observations. We find that strong electrostatic acceleration occurs on both sides of the termination shock, driven by magnetic reconnection. We also find that the dominant mode of particle escape changes from diffusive escape to advective escape as the blob passes through the shock.

\end{abstract}

\section{INTRODUCTION}

The Crab nebula is arguably the most well-studied calibration source for high-energy astrophysics due, to its proximity and the stability of its high-energy emission spectrum. It is a consistent source of electromagnetic radiation from radio to $\gamma$-rays, which is thought to be produced via synchrotron emission from electrons and positrons (hereafter, electrons) spiraling in the complex magnetosphere of the supernova remnant; see \citet{buehler14} for review. The pulsar, which powers the nebula, blows off a wind of relativistic electrons that propagates outwards with a bulk Lorentz factor on the order of $\Gamma \sim 10^6$ \citep{kennel84}. The magnetic field lines between the light cylinder and the termination shock are open and include both poloidal and toroidal components \citep{uzdensky14}, resulting in a ``cold," or radiationless zone. The plasma flows outward until the ram pressure balances the gas pressure, resulting in the formation of a termination shock at radius $r_t \sim 10^{17}\,$cm \citep{rees74,montani14}. Simulations indicate that the magnetic field in the vicinity of the termination shock is dominated by the toroidal component \citep{gallant92,uzdensky14,sironi15}.\\

Synchrotron radiation generated by relativistic electrons accelerated at the termination shock can explain the broad energy distribution of the observed quiescent emission \citep{gaensler06}. Electrons accelerated at the shock diffuse and advect outward, into the synchrotron-emitting outer region of the nebula located downstream from the termination shock \citep{kennel84,hester08}. As expected from the synchrotron process, the size of the nebula is inversely related to photon energy \citep{abdo11}, and therefore the most extended component is the radio emission, which illuminates the synchrotron nebula out to a radius of $\sim 10^{18}\,$cm. The Crab nebula is powered by the pulsar's spin-down luminosity, $\sim 5 \times 10^{38}\,\rm erg\,sec^{-1}$, which is transformed into electromagnetic radiation with an efficiency of $\sim 30$\% \citep{abdo11}.\\

Although the qualitative picture described above captures the essence of the mechanics of the Crab nebula's quiescent emission production, the details of the underlying particle acceleration processes are not understood very well. In particular, the standard model for first-order Fermi acceleration (also called diffusive shock acceleration, or DSA) at the termination shock does not seem to be able to explain the shape of the electron energy distribution implied by the observed quiescent synchrotron spectra, due to two complications \citep{komissarov13,olmi15}. The first is that the shock is relativistic instead of classical, since the upstream bulk Lorentz factor is on the order $\Gamma \sim 10^6$ \citep{lyubarsky03,aharonian04}. Somewhat paradoxically, relativistic shocks are less efficient accelerators than classical shocks \citep{ellison90}. Second, the magnetic field topology is probably toroidal in the vicinity of the termination shock. In this situation, electrons are less likely to be recycled back to the upstream side of the shock, and this further reduces the efficiency of the shock acceleration mechanism \citep{gallant92,sironi15}.\\

The theoretical challenges become much more severe when one turns attention from the quiescent emission to the series of five remarkable $\gamma$-ray flares observed by \textit{Fermi}-LAT starting in 2009 (along with an AGILE observation in 2007 September). The flares were characterized by about a ten-fold flux increase in the $\sim 0.1-1\,$GeV energy range, which is far beyond the cutoff in the quiescent spectrum at around $100\,$MeV. The brightest flare occurred in 2011 April, during which the Crab was the most luminous $\gamma$-ray source in the sky; hence this event is sometimes referred to as the ``super-flare'' \citep{striani11}. The 2011 April flare had a duration of about 9 days and consisted of two bright sub-flares, that each lasted $\sim 3-5$ days and displayed variability on sub-day timescales \citep{buehler12}.\\

The observed GeV emission from the Crab nebula suggests that extreme, impulsive particle acceleration is occurring near the termination shock on timescales of a few days. The large bulk Lorentz factor of the upstream nebular wind suggests that sufficient power is available to explain the observed $\gamma$-ray emission, if a suitably efficient acceleration mechanism can be identified \citep{buehler14}. The classical DSA mechanism is mediated by MHD waves, and therefore this process is limited to the Bohm rate \citep{lemoine09}. This results in a maximum photon energy, $\epsilon_{_{\rm MHD}}$, defined as the ``synchrotron burnoff'' limit \citep{cerutti13}, which is computed by equating the synchrotron loss timescale with the Larmor gyration timescale. The result obtained is \citep[e.g.,][]{kroon16}
\begin{equation}
\epsilon_{_{\rm MHD}} = \frac{6\pi q m_e c^2}{B_{\rm crit}\sigmaT} = 158 \, {\rm MeV}
\label{eq1}
\ ,
\end{equation}
where $m_e$ and $q$ denote the electron mass and charge, respectively, $c$ is the speed of light, $\sigmaT$ is the Thomson cross section, and $B_{\rm crit}=4.41 \times 10^{13}\,$G is the critical magnetic field. Equation~(\ref{eq1}) represents the classical upper limit on the synchrotron photon energy for electrons accelerated by any process that is mediated by MHD waves.\\

The problems with the classical DSA mechanism have motivated a variety of investigations into alternate acceleration mechanisms that may be operative at the termination shock, such as electrostatic acceleration via magnetic reconnection \citep[see][and references therein]{buehler14}. The classical synchrotron burnoff limit in Equation~(\ref{eq1}) results from the assumption that the acceleration process is mediated by MHD waves, and therefore this limit can be extended if one invokes electrostatic acceleration due to strong electric fields generated via magnetic reconnection \citep{cerutti12a,cerutti12b}. The induced electric fields efficiently accelerate the particles, while simultaneously reducing the magnetic field, thereby allowing the electrons to achieve very high Lorentz factors. The maximum photon energy can exceed the MHD photon energy limit in Equation (\ref{eq1}) due to the contribution from electrostatic acceleration, resulting in the new upper limit given by \citep{kroon16} 
\begin{equation}
\epsilon_{\textrm{max}} = 158 \, {\textrm{MeV}} \left(1 + \frac{E}{B} \right)
\label{eq1bbb}
\ ,
\end{equation}
where $E$ and $B$ denote the electric and magnetic fields, respectively. Once the electrons leave the acceleration region, they encounter a larger magnetic field, which converts their kinetic energy into a burst of high-energy synchrotron radiation. This process has been studied in detail using particle-in-cell (PIC) simulations \citep{cerutti13,cerutti14a,cerutti14b}, which demonstrated the feasibility and the general properties of this non-ideal MHD acceleration scenario. However, PIC simulations require long computation times, and therefore they are not amenable to fitting spectral data. This has motivated us to re-examine the problem using an analytical framework based on a robust, time-dependent particle transport equation that describes the evolution of the electron distribution during the observed $\gamma$-ray flares. The transport equation includes terms describing electrostatic acceleration as well as synchrotron losses and particle escape. Once an analytic solution to the transport equation is obtained, we can use it to compute the corresponding time-dependent synchrotron spectrum emitted by the relativistic electrons. The model can be used to approximately fit the $\gamma$-ray spectral data obtained using the \textit{Fermi}-LAT, while maintaining explicit control over the physical parameters.\\

\vspace{-0.3cm}

\section{Particle Transport Formalism}

\citet{kroon16} modeled the $\gamma$-ray spectrum observed from the Crab nebula during the peak of the 2011 April $\gamma$-ray flare using a steady-state model, based on the assumption that an approximate equilibrium prevails at the peak of the flare, implying that energy losses due to synchrotron emission are balanced by particle acceleration and particle escape. However, the event contained significant temporal structure, including two distinct sub-flares, which are clearly visible in the light curve plotted in Figure 5 from \citet{buehler12}. In order to understand the detailed temporal structure of this event, it is therefore necessary to develop a fully time-dependent model.\\ 

\begin{figure}[h!]
\vspace{0.0cm}
\centering
\includegraphics[height=9cm]{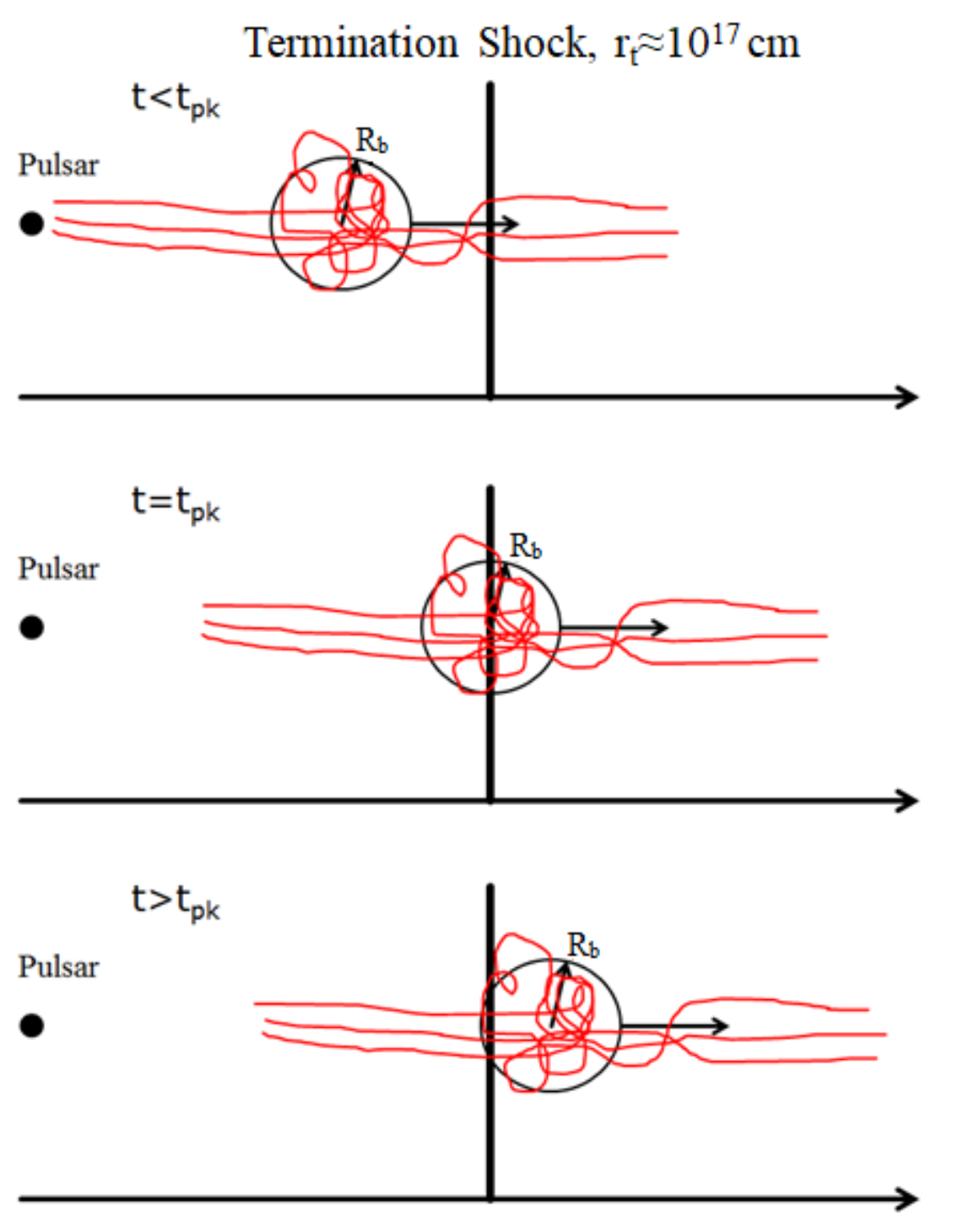}
\caption{This figure illustrates the physical picture in our model. The circles with radius $R_{\rm b}$ are a single blob shown at three different times and the red lines denote the magnetic fields. Each magnetically confined blob approaches the termination shock which causes shock-driven magnetic reconnection. The peak of a sub-flare corresponds to the middle image in which the blob is in maximum contact with the shock. The blob then advects downstream of the shock (lower image) where the escape mechanism switches from shock-regulated escape to advective particle escape.}
\label{fig1}
\end{figure}

We adopt the physical picture proposed by \citet{zrake16}, in which the relativistic electrons that produce the GeV $\gamma$-ray synchrotron emission observed during the flares are magnetically confined in a plasma ``blob'' that advects outward in the cold pulsar wind, eventually encountering the termination shock \citep[see also][]{uzdensky14}. Magnetic confinement imples that the radius of the blob, $R_{\textrm{b}}$, corresponds to the coherence length of the magnetic field, $\ell_{\textrm{coh}}$ \citep{kroon16}. We assume that the electrons in the blob have some initial momentum distribution, established in the upstream wind, and that the particles are further energized due to electrostatic acceleration when the blob passes through the shock. The electrons also lose energy via synchrotron radiation. We allow for the possibility of multiple plasma blobs, with each blob corresponding to a distinct sub-flare whose emission is part of the overall flare event. The blobs are assumed to be independent of each other, and they interact with the shock at different locations in time and space. In Figure \ref{fig1} we depict the interaction of a single blob with the shock.

\subsection{Particle Transport Equation}
\label{TransEqn}

The approach we take here is based on the electron transport equation analyzed by \citet{kroon16}, who found that the escape of electrons from the acceleration region was dominated by ``shock-regulated escape.'' In this scenario, relatively low-energy electrons with small Larmor radii are swept downstream from the termination shock and therefore escape from the acceleration region. Conversely, high-energy particles with large Larmor radii have a significant probability of scattering back into the upstream region, resulting in further acceleration in the vicinity of the termination shock. In this escape paradigm, the high-energy particles cycling back to the upstream region still remain inside the blob itself, because the effective mean-free path, $\ell$, is smaller than the magnetic coherence length, $\ell_{\textrm{coh}}$, which is essentially equal to the blob radius, $R_{\textrm{b}}$ \citep{kroon16}. We discuss further details of this in Section~\ref{parameterconstraints}. This physical scenario is applicable during the rising phase of each of the $\gamma$-ray sub-flares observed in 2011 April. However, as we argue below, the picture needs to be modified during the decaying phase of each sub-flare, during which the dominant escape mode for the relativistic electrons is probably better described by energy-independent advective escape rather than shock-regulated escape. Based on these considerations, in this paper we improve upon the steady-state model developed by \citet{kroon16} by solving two separate time-dependent electron transport equations, with one applicable during the rising phase of a sub-flare and the other applicable during the decaying phase.\\

We can model the evolution of the energy distribution of the blob electrons during the rising phase of a sub-flare using a transport equation that includes terms describing electrostatic acceleration, synchrotron losses, and particle escape. \citet{kroon16} found that momentum diffusion due to stochastic interactions with MHD waves does not contribute significantly to the acceleration of the particles during the observed $\gamma$-ray flares, and therefore electrostatic acceleration via shock-driven magnetic reconnection provides most of the flare's GeV luminosity. The electron transport equation describing this scenario can therefore be written as \citep{kroon16}
\begin{equation}
\frac{\partial f}{\partial t} = \frac{-1}{p^2}\frac{\partial}
{\partial p}\left\{p^2\left[A(t)m_e c -S(t) \frac{p^2}{m_e c} \right]f \right\}
- \frac{f}{t_{\rm esc}(p,t)}
\label{eq2}
\ ,
\end{equation}
where $t_{\rm esc}(p,t)$ denotes the timescale for particles to escape from the blob, which can in principle depend on both the time $t$ and the electron momentum $p$. The momentum distribution function for the electrons in the blob, $f(p,t)$, is related to the total number of electrons in the blob, $N_{\rm tot}$, via
\begin{equation}
N_{\rm tot}(t) = \int_0^\infty 4 \pi \, p^2\, f(p,t) \, dp \ .
\label{eq3}
\end{equation}
The time derivative on the left-hand side of Equation (\ref{eq2}) is interpreted as the Lagrangian rate of change in the frame of the plasma blob as it propagates outward through the cold pulsar wind in the region upstream from the termination shock.\\

In the context of our simplified one-zone model, we argue that the dominant form of particle escape, represented by the escape timescale, $t_{\rm esc}$, will differ qualitatively on the two sides of the termination shock. On the upstream side, we expect the particles to be governed by the so-called ``shock-regulated escape'' (SRE) process, in which high-energy particles are able to diffuse back to the upstream side of the shock as a result of their larger Larmor radii. The escape timescale $t_{\rm esc}$ is therefore energy-dependent on the upstream side of the shock. On the other hand, on the downstream side of the shock, particles tend to be swept away due to the compression of the flow, which traps the particles in the advecting, tangled magnetic field. In this situation, the escape timescale $t_{\rm esc}$ is expected to be energy-independent.\\

The functions $A(t)$ and $S(t)$ appearing in Equation~(\ref{eq2}) represent the time-dependent variation of the electrostatic acceleration and synchrotron loss terms, respectively. We can relate $A(t)$ and $S(t)$ to the time-dependent electric and magnetic fields, $E(t)$ and $B(t)$, respectively, by writing \citep{kroon16}
\begin{equation}
A(t) = \frac{q E(t)}{m_e c} \ , \qquad S(t) = \frac{\sigmaT B^2(t)}{6\pi m_e c} \ .
\label{eq4}
\end{equation}
We anticipate that the acceleration and loss processes experienced by the electrons in the blob will be strongly concentrated in the vicinity of the termination shock, where magnetic reconnection leads to strong electrostatic acceleration \citep{cerutti13}. In principle, the first order gain rate, $A(t)$, should also include a contribution due to shock acceleration; however, \citet{kroon16} demonstrated that electrostatic acceleration was at least an order of magnitude stronger than shock acceleration during the 2011 super-flare, which is the focus of the present paper. Hence, we will ignore shock acceleration here.\\
 
Following \citet{buehler12}, we assume that the physical variation of the magnetic and electric fields are correlated in time. Hence we introduce the ``profile function,'' $h(t)$, such that
\begin{equation}
A(t) = A_* h(t) \ , \qquad S(t) = S_* h(t) \ ,
\label{eq5}
\end{equation}
where the subscript ``$*$'' denotes the initial value of a quantity at the beginning of the $\gamma$-ray flare, at time $t=t_*$. The function $h(t)$ increases from the initial value $h(t_*)=1$, and reaches a maximum value during the peak of the flare, at time $t=t_{\rm pk}$, when the plasma blob is in maximum contact with the termination shock. The precise time dependence of the profile function, $h(t)$, is not known, but we can infer qualitative temporal trends by careful inspection of the time-domain data products of the flare. Based on the overall variation of the $\gamma$-ray light curves plotted in Figure 5 from \citet{buehler12}, we assume that the profile function, $h(t)$, increases exponentially during the rising portion of the sub-flare, and that it decreases exponentially during the decaying portion. This issue is discussed in more detail in Section~\ref{profilefunc}. We also note that we can combine Equations (\ref{eq4}) and (\ref{eq5}) to obtain the initial values of the functions $A(t)$ and $S(t)$, given by
\begin{equation}
A_* = \frac{q E_*}{m_e c} \ , \qquad S_* = \frac{\sigmaT B^2_*}{6\pi m_e c} \ ,
\label{eq6}
\end{equation}
where $E_*$ and $B_*$ denote the initial values of the electric and magnetic fields, respectively.\\

Combining Equations (\ref{eq4}) and (\ref{eq5}), we find that the time dependences of the electric and magnetic fields are given by
\begin{equation}
E(t) = E_* \, h(t) \ , \qquad B(t) = B_* \, \sqrt{h(t)} \ .
\label{eq7}
\end{equation}
Substituting Equations (\ref{eq5}) into Equation (\ref{eq2}) yields
\begin{equation}
\frac{\partial f}{\partial t} = \frac{-1}{x^2}\frac{\partial}
{\partial x}\left\{x^2\left[A_* h(t) - S_* h(t)x^2\right]f\right\}
- \frac{f}{t_{\rm esc}(x,t)}
\label{eq8}
\ ,
\end{equation}
where we have also introduced the dimensionless momentum, $x$, defined by
\begin{equation}
x \equiv \frac{p}{m_e c}
\label{eq9}
\ .
\end{equation}
Note that $x$ is related to the Lorentz factor, $\gamma$, via $x=\sqrt{\gamma^2-1}$. In our application to the $\gamma$-ray flares, we are focused on the evolution of a population of ultra-relativistic electrons, and therefore we can generally set $x=\gamma$.\\

It is preferable to work in terms of the electron number distribution, $N$, which is related to $f$ by
\begin{equation}
N(x,t) \equiv 4\pi (m_e c)^3 x^2f(x,t)
\label{eq14}
\ ,
\end{equation}
so that the total number of particles in the blob, $N_{\rm tot}$, is given by (cf. Equation~(\ref{eq3}))
\begin{equation}
N_{\rm tot}(t) = \int_0^\infty N(x,t) \, dx \ .
\label{eq15}
\end{equation}
By using Equation~(\ref{eq14}) to substitute for $f$ in Equation~(\ref{eq8}), we can rewrite the transport equation in the equivalent form
\begin{equation}
\frac{dN}{dt} = -\frac{\partial}
{\partial x}\left\{\left[A_* h(t) - S_* h(t)x^2 \right]N\right\}
- \frac{N}{t_{\rm esc}(x,t)} \ .
\label{eq16}
\end{equation}
It is convenient to non-dimensionalize the transport equation by dividing through by $A_* h(t)$, which yields
\begin{equation}
\frac{\partial N}{\partial y} = - \frac{\partial}
{\partial x}\left[\left(1- \hat S x^2\right)N\right]
- \frac{N}{A_* h(y) t_{\rm esc}(x,y)}
\label{eq10}
\ ,
\end{equation}
where
\begin{equation}
\hat S \equiv \frac{S_*}{A_*} \ ,
\label{eq11}
\end{equation}
and we have introduced the dimensionless time, $y$, defined by
\begin{equation}
dy \equiv A_* h(t)dt \ .
\label{eq12}
\end{equation}
The relationship between $y$ and $t$ is therefore given by the integral
\begin{equation}
y(t) = A_* \int_{t_*}^t h(t')dt'
\label{eq13}
\ ,
\end{equation}
where $A_*$ has units of ${\rm s}^{-1}$. Hence it follows that $y=0$ at the beginning of each sub-flare, at time $t=t_*$.

We note that Equation (\ref{eq10}) can also be expressed in the form of a Fokker-Planck equation by writing
\begin{equation}
\frac{\partial N}{\partial y} = \frac{\partial^2}{\partial x^2}
\left(\frac{1}{2} \, \frac{d\sigma^2}{dy} \, N\right)
- \frac{\partial}{\partial x}\left(\frac{dx}{dy}
\, N\right) - \frac{1}{A_* h(y) t_{\rm esc}(x,y)} \, N \ ,
\label{eq17}
\end{equation}
where the ``broadening'' and ``drift'' coefficients are given, respectively, by
\begin{equation}
\frac{1}{2} \, \frac{d\sigma^2}{dy} = 0 \ , \ \ \ \ \ 
\frac{dx}{dy} = 1-\hat S x^2
\ .
\label{eq18}
\end{equation}
The broadening coefficient vanishes in our application because the model under consideration here does not include momentum diffusion, which is negligible during the 2011 April super-flare \citep{kroon16}. Since there is no explicit source term in Equation (\ref{eq10}), in order to solve it, we must invoke an initial condition for the electron distribution at time $t=t_*$ ($y=0$), as discussed in Section~\ref{rpic}. We will solve Equation~(\ref{eq10}) to obtain the time-dependent electron energy distribution $N(x,y)$ during the rising and decaying phases of each $\gamma$-ray sub-flare in Sections~\ref{rpet} and \ref{dpet}, respectively.\\

\subsection{Single-Particle Evolution}

Since the broadening coefficient vanishes ($d\sigma^2/dt=0$), it follows that the transport equation is deterministic, and therefore it is possible to follow the evolution of the Lorentz factor of a single electron at any point in (dimensionless) time, $y$, given its initial Lorentz factor, $x_0$. We start with the differential equation (cf. Equation (\ref{eq18}))
\begin{equation}
\frac{dx}{dy} = 1-\hat S x^2 \ ,
\label{eq19}
\end{equation}
subject to the initial condition $x(0)=x_0$. Integration yields the solution
\begin{equation}
x(x_0,y) = \frac{1}{\sqrt{\hat S}} \, {\rm tanh}\left[{\rm tanh}^{-1}(x_0 \sqrt{\hat S})+y\sqrt{\hat S}\right]
\label{eq20}
\ .
\end{equation}
We note that as $y \to 0$, we recover the initial value $x \to x_0$ as required. On the other hand, in the limit $y \to \infty$, the value of $x$ approaches the equilibrium Lorentz factor, $\gamma_{\rm eq}$, defined by
\begin{equation}
\gamma_{\rm eq} \equiv \lim_{y \to \infty}x(x_0,y) = \frac{1}{\sqrt{\hat S}} \ .
\label{eq21}
\end{equation}
The electrons are consistently driven towards $\gamma_{\rm eq}$, regardless of their initial energy, because this is the Lorentz factor at which a balance is achieved between electrostatic acceleration and synchrotron losses. Hence we refer to $\gamma_{\rm eq}$ as the ``attractor energy.''\\

\begin{figure}[h!]
\vspace{0.0cm}
\centering
\includegraphics[height=6cm]{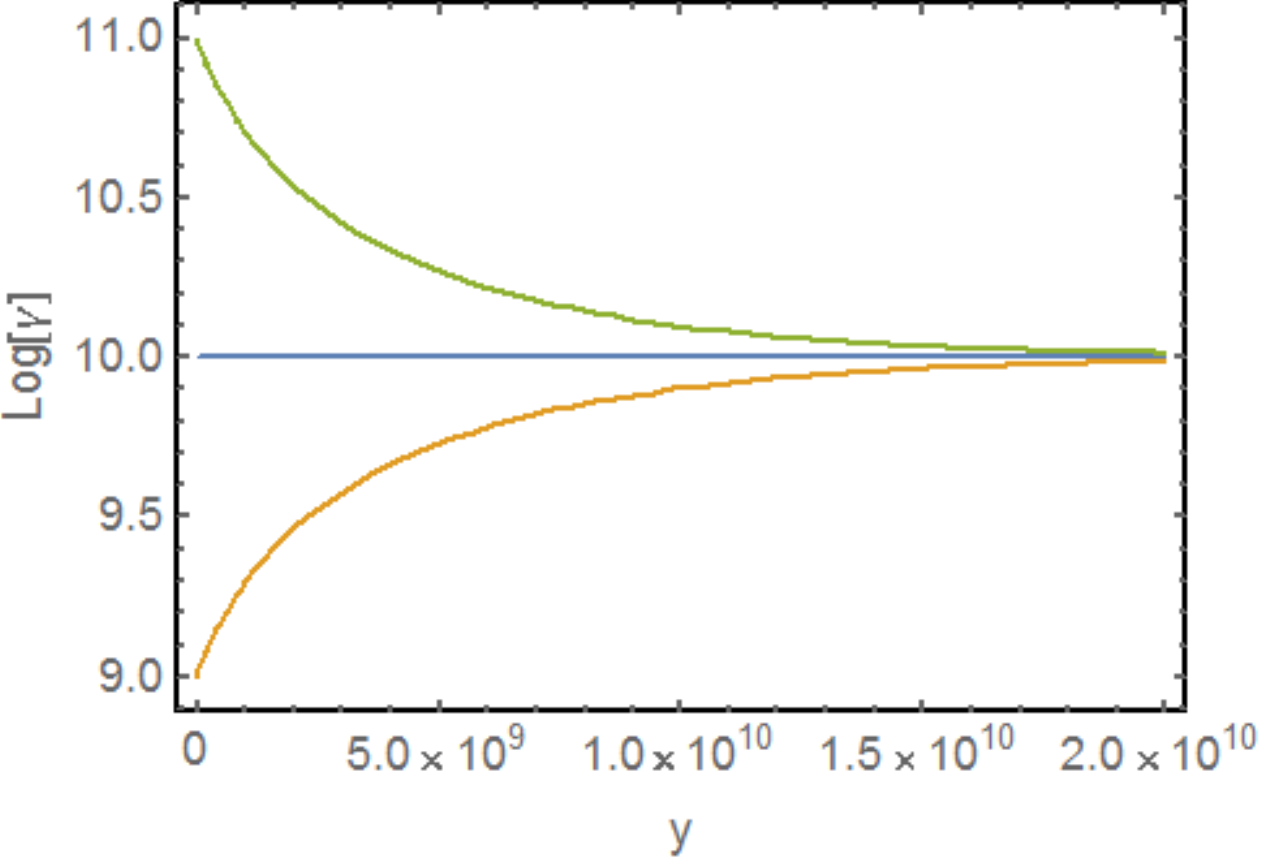}
\caption{The green curve represents the single-particle evolution of the energy $x$ when $x_0>\gamma_{\rm eq}$, given by Equation (\ref{eq20}). The orange curve denotes the single-particle evolution when $x_0<\gamma_{\rm eq}$. The blue line corresponds to the attractor energy, $\gamma_{\rm eq}$, given by Equation (\ref{eq21}).}
\label{fig2}
\end{figure}

\begin{figure}[h!]
\vspace{0.0cm}
\centering
\includegraphics[height=6cm]{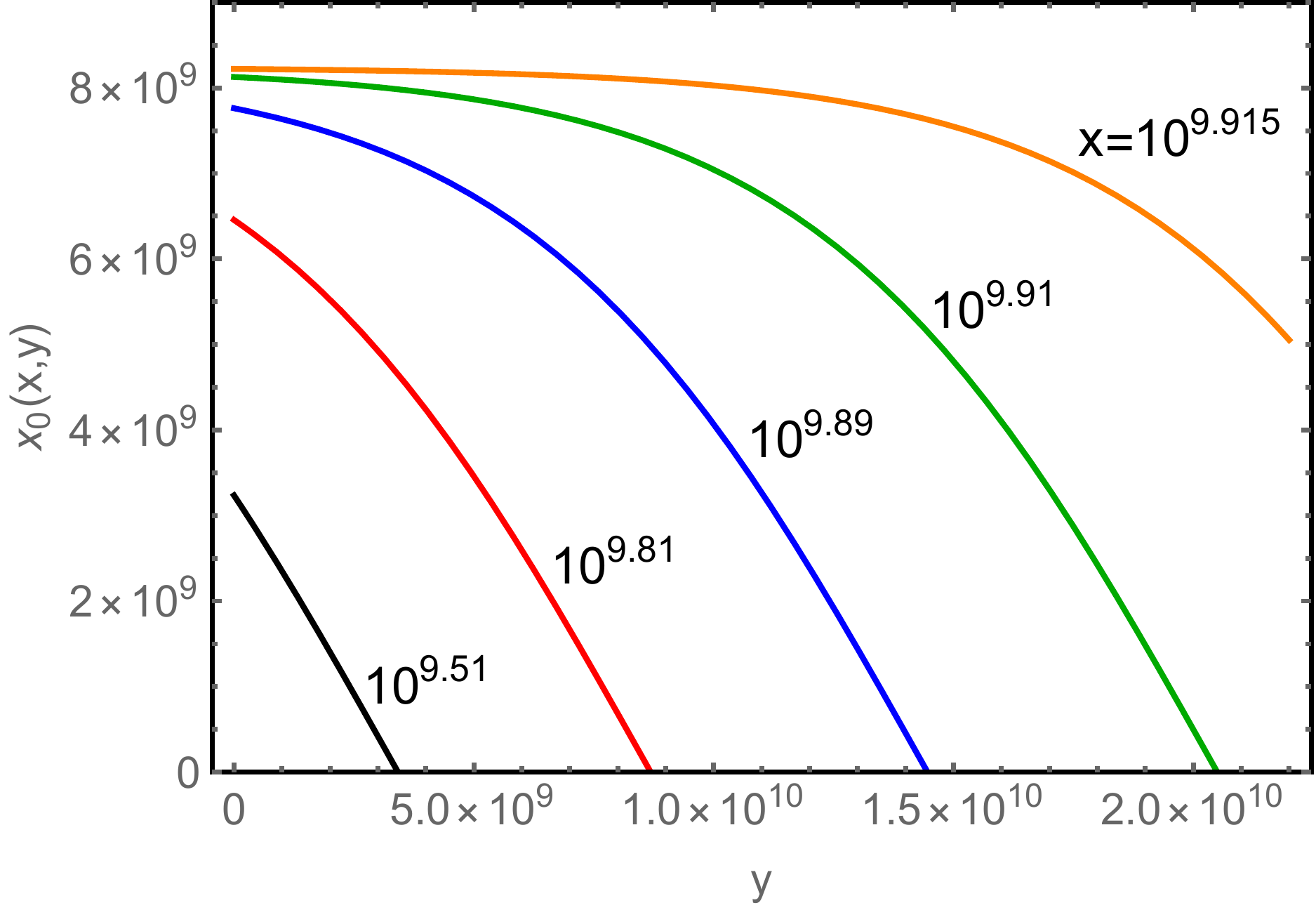}
\caption{This contour plot shows the initial energy of an electron, $x_0$ (Equation (\ref{eq23})) that currently has energy $x$ at dimensionless time $y$. The termination of each curve on the horizontal axis occurs at time $y=y_{\rm max}$ given by Equation (\ref{eq24}). In this example, we have set $\hat{S}=10^{-19.8}$.}
\label{fig2b}
\end{figure}

One can compute the lower limit for $x$ as a function of time $y$ by setting $x_0=0$ in Equation (\ref{eq20}), which yields
\begin{equation}
x_{\rm min}(y) = \frac{1}{\sqrt{\hat S}} \, {\rm tanh}\left(y\sqrt{\hat S}\right)
\label{eq22}
\ .
\end{equation}
This value corresponds to the Lorentz factor achieved at time $y$ by electrons initially injected with zero momentum. Hence at time $y$, all electrons must have Lorentz factors exceeding $x_{\rm min}(y)$. Equation (\ref{eq20}) can also be inverted to obtain the initial value of an electron's Lorentz factor, $x_0$, based on its current value, $x$, along with the current value of $y$. The result obtained is
\begin{equation}
x_0(x,y) = \frac{1}{\sqrt{\hat S}} \, {\rm tanh}\left[{\rm tanh}^{-1}(x \sqrt{\hat S})-y\sqrt{\hat S}\right]
\label{eq23}
\ .
\end{equation}
Since we must have $x_0 \ge 0$, Equation~(\ref{eq23}) implies that electrons with momentum $x$ cannot be observed after a maximum time $y_{\rm max}$, given by
\begin{equation}
y_{\rm max} = \frac{1}{\sqrt{\hat S}} \, {\rm tanh}^{-1}\left(x\sqrt{\hat S}\right)
\label{eq24}
\ .
\end{equation}
Equations (\ref{eq20}) and (\ref{eq21}) are plotted in Figure \ref{fig2}. Note that all of the electrons are driven towards the attractor energy, $\gamma_{\rm eq}$, as expected. In Figure \ref{fig2b} we plot Equation (\ref{eq23}) for several values of $x$ in order to demonstrate the effect of the evolving lower limit.

\subsection{Profile Function}
\label{profilefunc}

The profile function, $h(t)$, introduced in Equation (\ref{eq5}), represents the time-dependent modulation of the electric and magnetic fields due to impulsive reconnection occurring at the termination shock. As discussed above, the exponential shape of the $\gamma$-ray light curve during the 2011 April flare suggests that we adopt an exponential form for the profile function, which is consistent with models for magnetic reconnection \citep{buehler12}. This motivates the application of a piecewise exponential (rise/decay) function for the profile function. We therefore adopt the functional form
\begin{equation}
h(t) =
\begin{cases}
e^{\alpha \, t/t_{\rm pk}}, & {t \le t_{\rm pk}} \ , \\
e^{\alpha} e^{-\theta (\frac{t}{t_{\rm pk}}-1)}, & {t \ge t_{\rm pk}} \ ,
\end{cases}
\label{eq25}
\end{equation}
where $\alpha$ and $\theta$ are dimensionless constants, and $t_{\rm pk}$ is the time of peak intensity for a given sub-flare. \\

We can obtain an explicit expression for the variation of the dimensionless time, $y$, as a function of $t$ by combining Equations (\ref{eq13}) and (\ref{eq25}) and carrying out the integration. The result obtained is
\begin{equation}
y(t) =
\begin{cases}
\frac{A_* t_{\rm pk}}{\alpha}\left(e^{\alpha \, t/t_{\rm pk}}-1\right), & {t \le t_{\rm pk}} \ , \\
\frac{A_* t_{\rm pk}}{\alpha}\left(e^{\alpha}-1\right) - \frac{A_* t_{\rm pk}}{\theta}
e^{\alpha}\left[e^{-\theta(t/t_{\rm pk}-1)}-1\right], & {t \ge t_{\rm pk}} \ .
\end{cases}
\label{eq26}
\end{equation}
We find that unique solutions for the time-dependent sequence of $\gamma$-ray spectra (and the integrated light curves) can be found by varying the parameters $\alpha$ and $\theta$ along with the other theory parameters discussed below.\\

\section{Rising Phase Electron Transport}
\label{rpet}

In this study, we focus on the 2011 April Crab nebula $\gamma$-ray flare. This singular event is characterized by two ``sub-flares,'' which can be clearly seen as separate peaks in the $\gamma$-ray light curve plotted in Figure 5 from \citet{buehler12}. Each sub-flare peak is composed of a rising and decaying side. In order to compute the $\gamma$-ray synchrotron spectrum radiated during a given sub-flare, we must obtain the analytical solution for the energy distribution of the relativistic electrons in the plasma blob. In this section, we present the solution for the time-dependent electron distribution function applicable to the rising side of a single sub-flare component. This is obtained by solving the transport Equation (\ref{eq10}) subject to an initial condition discussed below.

\subsection{Rising Phase Electron Transport Equation}

Following \citet{kroon16}, we can model the rising phase evolution of the electron energy distribution in a plasma blob encountering and passing through the pulsar-wind termination shock using a transport equation that includes terms describing electrostatic acceleration, synchrotron losses, and shock-regulated escape (SRE). The corresponding transport equation is obtained by setting $t_{\rm esc}=t_{\rm SRE}$ in Equation (\ref{eq10}), where $t_{\rm SRE}$ is the mean escape time for the shock-regulated escape process, given by \citep{kroon16}
\begin{equation}
t_{\rm SRE}(p,t) = w(t) \, \frac{r_{\rm L}(p,t)}{c} \ , \qquad r_{\rm L}(p,t) = \frac{p \, c}{q B(t)} \ ,
\label{eq27a}
\end{equation}
where $r_{\rm L}$ is the Larmor radius, and $w$ denotes the efficiency factor for the SRE process, which accounts for the effects of time dilation and obliquity in the relativistic shock. Larger values of $w$ imply longer escape times (at a given particle energy), and therefore more efficient recycling of particles back to the upstream side of the shock. Hence larger values of $w$ tend to enhance the acceleration process by inhibiting the escape of particles from the shock. We can also write the SRE timescale in the form
\begin{equation}
t_{\rm SRE}(p,t) = \frac{p}{C(t) m_e c} \ , \qquad C(t) = C_* h(t) \ ,
\label{eq27}
\end{equation}
where the function $C(t)$ is defined in terms of physical quantities by writing
\begin{equation}
C(t) = \frac{q B(t)}{w(t)m_e c} \ .
\label{eq31}
\end{equation}

Combining Equations~(\ref{eq10}) and (\ref{eq27}) yields the transport equation for the rising phase of the sub-flare. The result obtained is
\begin{equation}
\frac{dN}{dy} = -\frac{\partial}
{\partial x}\left[\left(1- \hat S x^2\right)N\right]
- \frac{\hat C N}{x} \ ,
\label{eq29}
\end{equation}
where $N(x,y)$ is the electron number distribution defined in Equation~(\ref{eq14}), and we have defined the dimensionless constants
\begin{equation}
\hat S \equiv \frac{S_*}{A_*} \ , \qquad \hat C = \frac{C_*}{A_*} \ .
\label{eq30}
\end{equation}
The corresponding initial value of $C(t)$ is given by
\begin{equation}
C_* = \frac{q B_*}{w_*m_e c} \ ,
\label{eq32}
\end{equation}
where $w_*$ denotes the initial value of the SRE efficiency factor. By combining Equations~(\ref{eq27}), (\ref{eq31}), and (\ref{eq32}), we conclude that the variation of $w(t)$ is given by
\begin{equation}
w(t) = \frac{w_*}{\sqrt{h(t)}} \ .
\label{eq33}
\end{equation}
The physical significance of this variation will be further discussed below.

\subsection{Rising Phase Initial Condition}
\label{rpic}

In order to solve Equation (\ref{eq29}), we need to impose an initial condition for the electron distribution, $N(x,y)$, which describes the momentum distribution of the electrons in the cold striped wind. We assume the initial distribution to be a Gaussian given by
\begin{equation}
N(x,y)\Big|_{y=0} = \frac{J_0}{\sigma \sqrt{2\pi}}e^{\frac{-(x-\mu)^2}{2\sigma^2}} \ ,
\qquad \frac{1}{\sqrt{\hat S}} \ge x \ge 0 \ ,
\label{eq34}
\end{equation}
where $\mu$ and $\sigma$ denote the mean and standard deviation, respectively, and $J_0$ is the normalization factor. The electron Lorentz factor has an upper bound at $\gamma_{\rm eq}=\hat{S}^{-1/2}$ (c.f. Equation (\ref{eq21})) due to the fact that synchrotron losses overwhelm electrostatic acceleration beyond this energy. Furthermore, electrons with an initial energy above this upper limit would readily lose energy until reaching the attractor Lorentz factor. Thus, $\hat{S}^{-1/2}$ represents the upper limit for the electron Lorentz factor throughout this study.\\

The dimensionless momentum $x$ cannot be less than zero, which corresponds to a minimum Lorentz factor of 1. Therefore, for a given value of $J_0$, the total number of electrons, $\mathscr{N}_0$, initially contained in the blob is computed using
\begin{equation}
\mathscr{N}_0 \equiv N_{\rm tot}(t_{*}) = J_0 \int_{0}^{1/\sqrt{\hat{S}}}\frac{1}{\sigma\sqrt{2\pi}}e^{\frac{-(x-\mu)^2}{2\sigma^2}}dx
= \frac{J_0}{2}\left[{\rm Erf}\left(\frac{\mu}{\sigma\sqrt{2}}\right)-{\rm Erf}\left(\frac{\mu-1/\sqrt{\hat{S}}}{\sigma\sqrt{2}}\right)\right]
\label{eq35}
\ .
\end{equation}
The total energy in the initial electron distribution, $\mathscr{E}_0$, is likewise given by
\begin{equation}
\mathscr{E}_0 = J_0 \int_{0}^{1/\sqrt{\hat S}}\frac{m_e c^2\sqrt{x^2+1}}{\sigma\sqrt{2\pi}}e^{\frac{-(x-\mu)^2}{2\sigma^2}}dx
\label{eq36}
\ ,
\end{equation}
since $\gamma=\sqrt{x^2+1}$ is the general expression for the Lorentz factor. Hence, the mean value for the Lorentz factor of the initial electron distribution in the blob is given by
\begin{equation}
\bar \gamma_0 = \sqrt{\bar{x}^2_0+1} = \frac{\mathscr{E}_0}{\mathscr{N}_0m_e c^2}
\label{eq37}
\ .
\end{equation}
We find that  in our application to the Crab nebula $\gamma$-ray flares, $\bar{\gamma}_0 \sim 10^9$.\\ 

The fact that $\bar{\gamma}_0$ greatly exceeds the upstream Lorentz factor in the cold pulsar wind, $\Gamma \sim 10^6$, suggests that the electrons in the blob represent a population produced as a result of impulsive reconnection in the region just upstream from the termination shock \citep{cerutti14a}. The electrons are pre-accelerated (due to explosive reconnection) at a faster rate than the synchrotron cooling timescale, and therefore very little synchrotron emission is produced during this phase. By contrast, the timescale for the subsequent acceleration is comparable to the synchrotron timescale, and therefore most of the flare emission is produced on timescales of a few days.


\subsection{Rising Phase Electron Distribution}

We can obtain the exact solution for the electron distribution in the plasma blob during the rising portion of the sub-flare by solving the transport Equation (\ref{eq29}) subject to the initial condition given by Equation (\ref{eq34}). The exact solution for the electron distribution during the rising phase of a sub-flare is given by
\begin{equation}
N_{\rm rise}(x,y) = \frac{J_0}{\sigma \sqrt{2\pi}}
\left[\frac{x_0(x,y)}{x}\right]^{\hat C}
\left[\frac{1-\hat S x_0^2(x,y)}{1-\hat S x^2}\right]^{1-\frac{\hat C}{2}}
{\rm exp}\left\{-\frac{[\mu + x_0(x,y)]^2}{2\sigma^2}\right\} \ \ , \ \ x_{\rm min}(y)< x<\frac{1}{\sqrt{\hat{S}}},
\label{eq38}
\end{equation}
where $x_{\rm min}(y)$ is computed using Equation (\ref{eq22}), $x_0(x,y)$ is evaluated using Equation~(\ref{eq23}), and $y$ is evaluated as a function of time $t$ using Equation~(\ref{eq26}). This solution for the particle distribution function applies during the rising portion of each sub-flare, $t \le t_{\rm pk}$, or $y \le y_{\rm pk}$.\\ 

\section{Decaying Phase Electron Transport}
\label{dpet}

In this section we present the derivation of the time-dependent electron distribution function describing the particle transport occurring during the decaying phase of each sub-flare. The escape of particles in the decaying phase is dominated by advection, because after the peak of the sub-flare, the blob has moved downstream from the termination shock. Thus, the shock-regulated escape mechanism is inapplicable to the decaying phase of the sub-flare evolution.\\

\subsection{Decaying Phase Electron Transport Equation}

In Section~\ref{rpet}, we presented the solution for the transport equation during the rising phase of the sub-flare, which is characterized by electrostatic acceleration, synchrotron losses, and shock-regulated escape. This escape mechanism is valid while the blob is approaching and moving through the shock. However, after most of the blob has moved downstream from the shock, the situation changes, because magnetic confinement will sweep all of the particles downstream at the same rate, independent of their energy (see Figure \ref{fig1}). Thus, we adopt an energy-independent, advective escape paradigm during the decaying phase of each sub-flare. Furthermore, we assume that the escape timescale, $t_{\rm esc}$, is independent of time, reflecting particle advection occurring at a rate corresponding to the downstream velocity of the flow leaving the shock, which is $c/3$ \citep{achterberg01}.\\ 

As the plasma blob advects downstream from the termination shock, electrostatic acceleration continues due to shock-induced magnetic reconnection, since relatively strong electrostatic fields can persist downstream from the shock for a couple of light-days in the Crab nebula \citep{cerutti13, cerutti14a, cerutti14b}. Compression of the flow leads to an increase in the likelihood that the particles will be swept away in the tangled magnetic field, making advective escape the dominant channel for particles to leave the vicinity of the termination shock. In this scenario, the escape timescale is independent of both energy and time, and we can therefore write $t_{\rm esc}=t_{\rm ad}$ in Equation (\ref{eq16}), where $t_{\rm ad}$ is the advective timescale. The advective timescale can be estimated by writing
\begin{equation}
t_{\rm ad} = \frac{R_{\rm b}}{v_{ds}} \ ,
\label{eq40}
\end{equation}
where $R_{\rm b}$ is the blob radius and $v_{ds}=c/3$ denotes the downstream flow velocity \citep{achterberg01}. The resulting transport equation applicable during the decaying phase of the sub-flare is
\begin{equation}
\frac{\partial N}{\partial t} = - \frac{\partial}{\partial x}
\left\{\left[A_* h(t) - S_* h(t)x^2 \right]N\right\} - \frac{N}{t_{\rm ad}} \ .
\label{eq41}
\end{equation}
Since the escape timescale $t_{\rm ad}$ is assumed to be independent of energy and time, it follows that the solution for $N$ can be written in the form
\begin{equation}
N(x,y) = e^{-(t-t_{\rm pk})/t_{\rm ad}}G(x,y) \ ,
\label{eq42}
\end{equation}
where $t_{\rm pk}$ denotes the peak time for the sub-flare, and $G(x,t)$ is the solution to the equation
\begin{equation}
\frac{dG}{dt} = -\frac{\partial}
{\partial x}\left\{\left[A_* h(t) - S_* h(t)x^2 \right]G\right\}
\label{eq43}
\ .
\end{equation}
By transforming to the dimensionless time $dy=A_* h(t) dt$, we can obtain the equivalent expression
\begin{equation}
\frac{dG}{dy} = - \frac{\partial}
{\partial x}\left[\left(1- \hat S x^2\right)G\right] \ ,
\label{eq44}
\end{equation}
where we have defined the dimensionless constant
\begin{equation}
\hat S \equiv \frac{S_*}{A_*} \ .
\label{eq45}
\end{equation}
Once Equation~(\ref{eq44}) has been solved to determine the function $G(x,y)$, we can obtain the solution for the electron momentum distribution $N(x,y)$ by employing Equation~(\ref{eq42}).

\subsection{Decaying Phase Initial Condition}

In order to solve Equation (\ref{eq44}) to determine the electron distribution during the decaying phase of the sub-flare, we need to impose an initial condition for the electron distribution, $N(x,y)$, at the initial time, which in this case is $y=y_{\rm pk}$, or $t=t_{\rm pk}$. The appropriate initial condition for the decaying phase of the sub-flare is provided by evaluating the rising-phase solution, $N_{\rm rise}(x,y)$, given by Equation~(\ref{eq38}), at the peak of the sub-flare, $t=t_{\rm pk}$. The initial condition for the decaying phase is therefore given by
\begin{equation}
N_{\rm decay}(x,y_{\rm pk}) = N_{\rm rise}(x,y_{\rm pk}) \ ,
\label{eq46}
\ 
\end{equation} 
where $N_{\rm rise}(x,y_{\rm pk})$ is evaluated using Equation (\ref{eq38}). Equation~(\ref{eq46}) also ensures that the electron distribution $N(x,y)$ is continuous across the peak of the sub-flare.

The exact solution to Equation (\ref{eq41}) subject to this initial condition can be written as
\begin{multline}
N_{\rm decay}(x,y) = \frac{J_0 \, e^{-(t-t_{\rm pk})/t_{\rm ad}}}
{\sigma \sqrt{2\pi}}
\left[\frac{1-\hat S x_0^2(x,y)}{1-\hat S x^2}\right]
{\rm exp}\left\{-\frac{\left[\mu+x_0(x,y)\right]^{2}}{2\sigma^2}\right\}
\\
\times \left[\frac{1-\hat S x_0^2(x,y-y_{\rm pk})}{1-\hat S x_0^2(x,y)}\right]^{\hat C/2}
\left[\frac{x_0(x,y-y_{\rm pk})}{x_0(x,y)}\right]^{-\hat C}\ , \qquad x_{\rm min}(y)< x<\frac{1}{\sqrt{\hat{S}}},\, \ \ y \ge y_{\rm pk},
\label{eq47}
\end{multline}
where $x_0(x,y)$ and $x_0(x,y-y_{\rm pk})$ are evaluated using Equation~(\ref{eq23}). The results we have obtained in Equations (\ref{eq38}) and (\ref{eq47}), respectively, represent the exact solutions for the electron energy distribution during the rising and decaying phases of a single sub-flare.

\section{Synchrotron Spectra}

Since synchrotron losses are included in the transport equations we have used to treat both the rising and decaying phases of the sub-flare, we can therefore compute self-consistent synchrotron spectra and compare those results with the $\gamma$-ray spectra observed during the 2011 April super-flare. Assuming an isotropic distribution of electrons, the theoretical synchrotron spectrum can be computed by convolving the electron number distribution function, Equations (\ref{eq38}) and (\ref{eq47}), with the single-particle synchrotron emission function, $Q_\nu$, given by \citep[e.g.,][]{becker92,kroon16}
\begin{equation}
Q_\nu(\nu,\gamma) = \frac{\sqrt{3}\,q^3 B}{m_e c^2}R
\left(\frac{\nu}{\gamma^2 \nu_s}\right) \ \ \propto \ \ {\rm erg \ s^{-1}
\ Hz^{-1}}
\ ,
\label{eq48}
\end{equation}
where
\begin{equation}
\nu_s \equiv \frac{3 q B}{4 \pi m_e c}
\ ,
\label{eq49}
\end{equation}
and \citep{crusius86}
\begin{equation}
R(x) \equiv
\frac{x^2}{2}K_{4/3}\Big(\frac{x}{2}\Big)K_{1/3}\Big(\frac{x}{2}\Big)-
\frac{3x^3}{20}\Big[K^2_{4/3}\Big(\frac{x}{2}\Big)-
K^2_{1/3}\Big(\frac{x}{2}\Big)\Big]
\ .
\label{eq50}
\end{equation}
Here, $K_{4/3}(x)$ and $K_{1/3}(x)$ denote modified Bessel functions of the second kind.
The synchrotron spectrum emitted by the electron distribution in the co-moving frame of the blob is 
computed by performing the integral convolution
\begin{equation}
P_\nu(\nu,t) = \int_{x_{\rm min}[y(t)]}^\infty N[x,y(t)]
Q_\nu(\nu,x)dx \ \propto \ {\rm erg \ s^{-1} \ Hz^{-1}}
\ ,
\label{eq51}
\end{equation}
where $x_{\rm min}$ is computed using Equation (\ref{eq22}). Although Equation (\ref{eq51}) technically applies in the frame of the blob, the transformation to the observer's frame is insignificant because the flow velocity downstream from the shock is equal to $c/3$, with a bulk Lorentz factor $\Gamma=1.06$ \citep{achterberg01}. The electron distribution $N(x,y)$ is evaluated using either $N_{\rm rise}(x,y)$ (Equation~(\ref{eq38})) during the rising phase of the sub-flare and using $N_{\rm decay}(x,y)$ (Equation (\ref{eq47})) during the decaying phase. Lastly, we note that integrating over the dimensionless momentum, $x$, or Lorentz factor, $\gamma$, are both approximately equal since the electrons producing the flare emission have Lorentz factors $\gamma \gg 1$. The corresponding theoretical flux levels are given by
\begin{equation}
\mathscr{F}_\nu(\nu,t) = \frac{P_\nu(\nu,t)}{4 \pi D^2}
\ \propto \ {\rm erg \ s^{-1} \ cm^{-2} \ Hz^{-1}}
\ ,
\label{eq52}
\end{equation}
where $D$ is the distance to the source and $P_\nu(\nu,t)$ is computed using Equation (\ref{eq51}).\\

The spectra plotted by \citet{buehler12} include the background component due to the synchrotron nebula, which is a steep power-law whose functional form can be inferred from their plots. We have found the time-independent background flux, $\mathscr{F}^{\rm neb}$, to be
\begin{equation}
\mathscr{F}_{\nu}^{\rm neb}(\nu)=1.18 \times 10^{35} \left(\frac{\nu}{\rm Hz}\right)^{-3} \ \propto \ {\rm erg \ s^{-1} \ cm^{-2} \ Hz^{-1}}
\label{eq53}
\ .
\end{equation}
We will integrate this background flux over the appropriate photon energy range ($0.1-100\,$GeV) to compute the background integrated flux for the light curve.\\

\section{Application to 2011 April Flare}

In this section we present a comparison of the theoretically computed spectral snapshots and integrated light curve with observational data of the 2011 April super-flare. We remind the reader that a sub-flare is composed of rising and decaying phases each of which is modeled using the appropriate solution for the particle distribution function.\\ 

\subsection{Parameter Calculations}

We have thus far presented a detailed derivation of the time-dependent electron number distribution for the rising and decaying phases of a sub-flare given by Equations (\ref{eq38}) and (\ref{eq47}), respectively. We can port these solutions for $N(\gamma,y)$ into Equations (\ref{eq51}) and (\ref{eq52}) to compute the observed photon energy spectrum $\mathscr{F}_\nu(\nu,t)$ at any point in time during a sub-flare.\\ 

For a given sub-flare, the application of our model requires the specification of values for the ten free parameters $J_0$, $\mu$, $\sigma$, $\hat S$, $\hat C$, $E_*/B_*$, $t_*$, $t_{\rm ad}$, $\alpha$, and $\theta$. We remind the reader that the parameters $J_0$, $\mu$, $\sigma$ describe the number of electrons initially in the blob, the Gaussian mean, and the standard deviation, respectively. The parameters $\hat S$, $\hat C$, $E_*/B_*$, and $t_{\rm ad}$ describe the dimensionless synchrotron loss rate, the dimensionless shock-regulated escape rate, the ratio of electric to magnetic fields at MJD date $t_*$ (the temporal origin of the sub-flare), and the advective escape timescale, respectively. Lastly, the parameters $\alpha$ and $\theta$ characterize the profile function, $h(t)$, (see Section~\ref{profilefunc}). A quantity that is needed to characterize the profile function is $t_{\rm pk}$, which is the time in seconds from the initial time $t_*$ to the peak of the sub-flare. The date of each peak is not arbitrary, but is simply found by visual inspection of the light curve. The initial magnetic field, $B_*$, and electric field, $E_*$, can be computed by dividing Equation~(\ref{eq6}) for $S_*$ by $A_*$ and then solving for $B_*$, which yields
\begin{equation}
B_* = \frac{6\pi q \hat S}{\sigmaT}\left(\frac{E_*}{B_*}\right) \ ,
\qquad E_* = \frac{6\pi q \hat S}{\sigmaT}\left(\frac{E_*}{B_*}\right)^2
\label{eq54}
\ ,
\end{equation}
where the quantities $\hat S$ and $(E_*/B_*)$ are free parameters in our model. The corresponding values of $A_*$ and $C_*$ obtained by applying Equations~(\ref{eq6}) and (\ref{eq30}) are given by
\begin{equation}
A_* = \frac{6\pi q^2 \hat S}{\sigmaT m_e c}\left(\frac{E_*}{B_*}\right)^2 \ ,
\qquad C_* = \frac{6\pi q^2 \hat S \hat C}{\sigmaT m_e c}\left(\frac{E_*}{B_*}\right)^2
\label{eq54b}
\ ,
\end{equation}
where $\hat C$ is also a model free parameter. Finally, the initial value for the SRE efficiency parameter, $w_*$, is obtained by combining Equations~(\ref{eq32}) and (\ref{eq54b}), which yields
\begin{equation}
w_* = \frac{1}{\hat C}\left(\frac{E_*}{B_*}\right)^{-1} \ .
\label{eq54c}
\end{equation}
The relations discussed above allow us to compute all necessary physical parameters in terms of the model free parameters.

\subsection{Spectra}

\begin{figure}[h!]
\vspace{0.0cm}
\centering
\includegraphics[height=11cm]{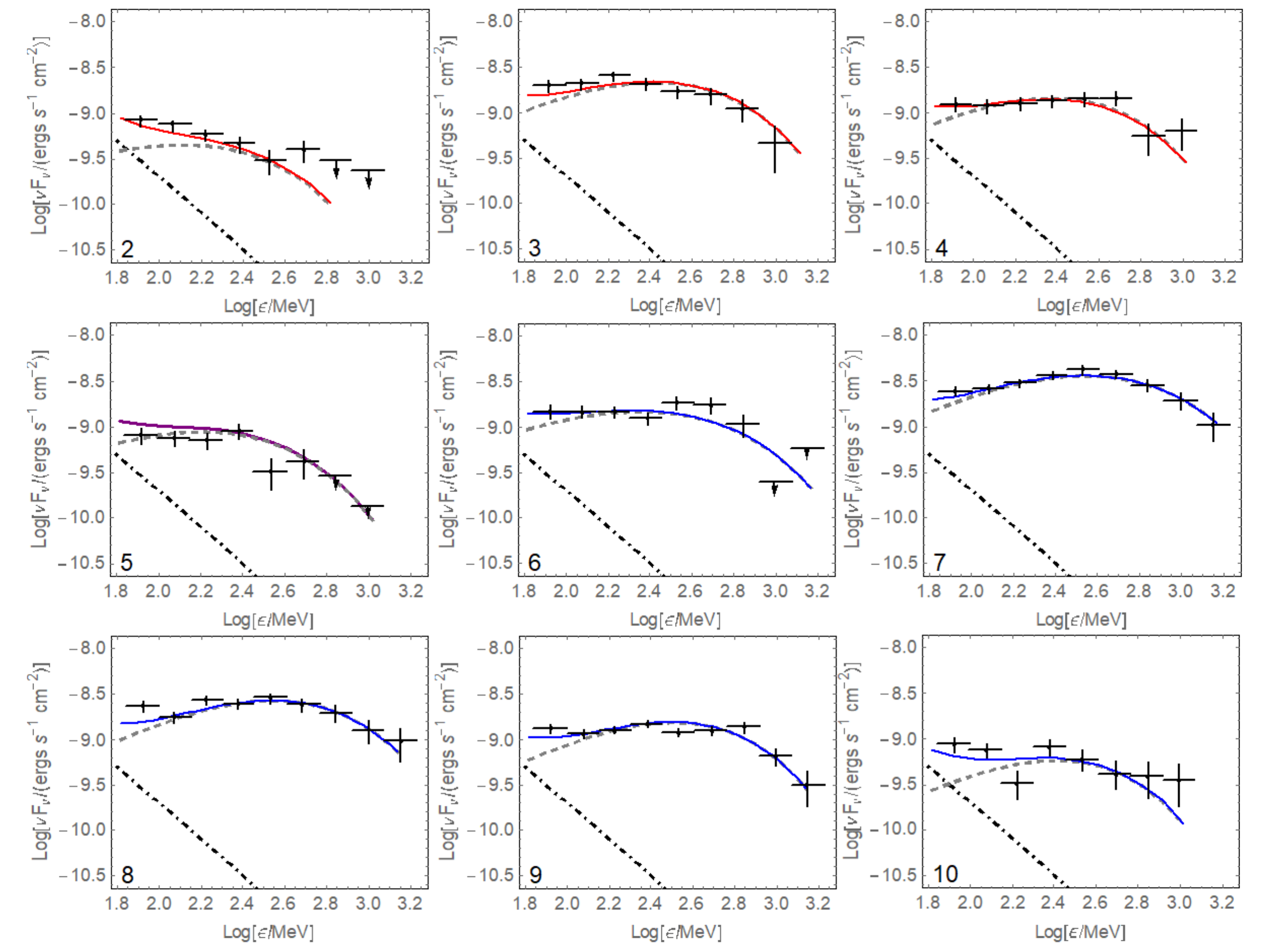}
\caption{Theoretical time-dependent $\gamma$-ray flare spectrum computed using Equation (\ref{eq52}), plotted as a function of the photon energy (gray dashed curve). The solid colored curves represent the sum of the theoretical component and the background nebular spectrum (dot-dashed curve). The theory curves include the contributions from both plasma blobs. The red curves in panels 2-4 indicate that the emission is dominated by Blob 1, and the blue curves in panels 6-10 indicate that the emission is dominated by Blob 2. The purple curve in panel 5 indicates that both blobs contribute about equally to the theoretical spectrum. The spectral data are from \citet{buehler12}, and the index numbers in the lower left-hand corners correspond to their time interval notation.}
\label{fig3}
\end{figure}

In order to test the model under study here, we need to compare our predictions with the observational data presented by \citet{buehler12}. Specifically, in their Figure 5, the evolving flare spectrum is depicted in a series of panels whose time indices correspond to the numbered time intervals indicated in the light curve. They present spectral curves for the flaring component as well as the background nebula emission (constant in time). In Figure \ref{fig3}, we compare our spectral curves, $\mathscr{F}_{\nu}(\nu,t)$, computed using Equation (\ref{eq52}), with the corresponding observational data reported by \citet{buehler12}. Each panel includes a time index number in the lower left-hand corner that corresponds to the time indices in the light curve plotted by \citet{buehler12}.\\

The theoretical spectra plotted in Figure \ref{fig3} were computed by summing the individual synchrotron contributions from the two blobs at each point in time. It is interesting to note that the relative contributions of the two plasmas blobs change during the course of the super-flare. From time intervals 2-4, the synchrotron emission from Blob 1 dominates the composite spectrum, and from time intervals 6-10, the emission from Blob 2 dominates. The peak of the first sub-flare occurs at time interval 3. Hence the spectra in time intervals 2 and 4 were computed using the electron distributions $N_{\rm rise}$ and $N_{\rm decay}$, respectively, for Blob 1. The spectrum in time interval 3 can be computed using either electron distribution, since the electron distribution is continuous at the peak of the sub-flare (see Equation~(\ref{eq46})). Note that time interval 5 is unique in the sense that at this particular time, each sub-flare (and therefore each of the two blobs) contributes about equally to the observed emission. Panel 5 therefore represents a transition between the two sub-flares, and the spectrum is given by the sum of the decaying phase emission from Blob 1 and the rising phase emission from the Blob 2.\\

The peak of the second sub-flare occurs at time interval 7, and therefore the spectra in time intervals 5-7 were computed using the electron distribution $N_{\rm rise}$ for Blob 2, whereas the spectra in time intervals 7-10 were computed using the electron distribution $N_{\rm decay}$ for Blob 2. Time interval 7 can be computed using either electron distribution, since the distribution is continuous at the peak of the sub-flare according to Equation~(\ref{eq46}). We note that overall, the time-dependent $\gamma$-ray spectrum for the 2011 April super-flare computed using our model and plotted in Figure \ref{fig3} agrees quite well with the data from \citet{buehler12} for all of the time intervals. This agreement supports the electrostatic acceleration framework explored here.\\

\subsection{Electron Distributions}

\begin{figure}[h!]
\vspace{0.0cm}
\centering
\includegraphics[height=7.5cm]{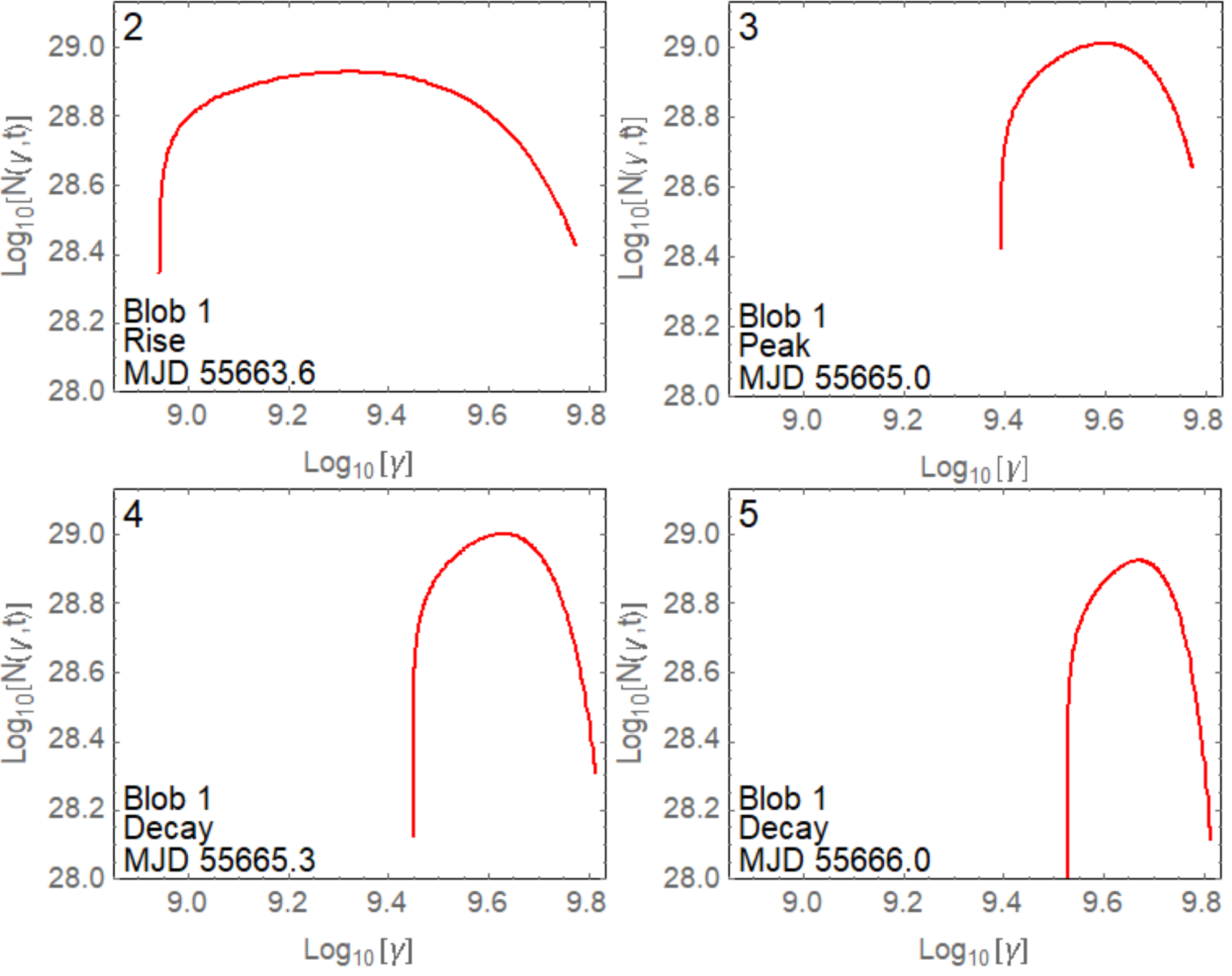}
\caption{Electron distribution function, $N(\gamma,t)$, for the rising and decaying phases of Blob 1. Panels 2 and 3 are plotted using Equation (\ref{eq38}) and panels 4 and 5 are plotted using Equation (\ref{eq47}). The index numbers in the upper left-hand corners correspond to the time intervals discussed by \citet{buehler12}.}
\label{fig4}
\end{figure}

\begin{figure}[h!]
\vspace{0.0cm}
\centering
\includegraphics[height=7.5cm]{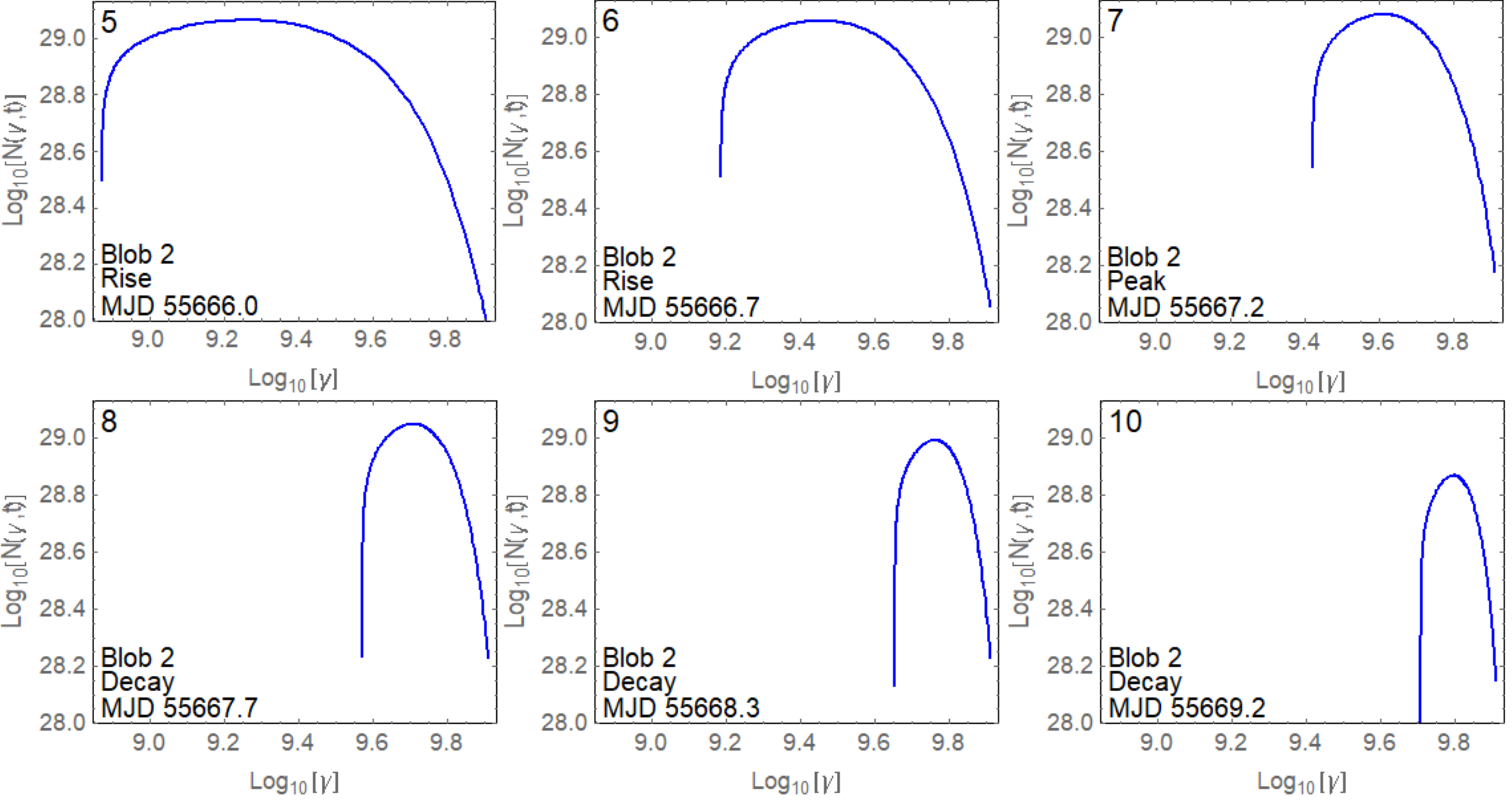}
\caption{Same as Figure \ref{fig4}, except the electron distribution function, $N(\gamma,t)$, corresponds to the rising and decaying phases of Blob 2. Panels 5, 6, and 7 are plotted using Equation (\ref{eq38}) and panels 8, 9, and 10 are plotted using Equation (\ref{eq47}).}
\label{fig5}
\end{figure}

The $\gamma$-ray spectra plotted in Figure \ref{fig3} were computed using either the rising- or decaying-phase solutions for the electron distribution function, $N_{\rm rise}$ and $N_{\rm decay}$, given by Equations (\ref{eq38}) and (\ref{eq47}), respectively. In Figures \ref{fig4} and \ref{fig5}, we plot the corresponding electron distributions in each time window during the flare. The electron distribution functions, regardless of phase, exhibit a low and high-energy cutoff. The high-energy cutoff corresponds to the attractor energy, $\gamma_{\rm eq}$, which represents a balance between electrostatic acceleration and synchrotron losses (see Equation (\ref{eq21})). The electron distributions also exhibit a low-energy cutoff, at $x_{\rm min}$ (Equation (\ref{eq22})), corresponding to the current energy of electrons that were injected with zero energy at the beginning of the sub-flare, at time $t=t_*$. This effect causes the low-energy cutoff seen in the electron distributions in Figures \ref{fig4} and \ref{fig5} to drift to higher energies with increasing time. The electron distribution becomes more narrow as the electrons get ``squeezed'' between the evolving low-energy cutoff $x_{\rm min}$ and the fixed attractor energy, above which the electrons cannot be accelerated. The amplitude of the distributions also decreases with time due to the escape of electrons from the plasma blob.\\

Each panel in Figures \ref{fig4} and \ref{fig5} includes a time index number in the upper left-hand corner that corresponds to the time intervals for the $\gamma$-ray spectra plotted in Figure~\ref{fig3}. Figure \ref{fig4} depicts the electron distribution functions during the first sub-flare, and thus represent the evolving momentum distribution of the electrons in Blob 1. Figure \ref{fig5} depicts the evolving momentum distribution for the electrons in Blob 2, corresponding to the second sub-flare. We note that the momentum distribution at the peak of a given sub-flare can be computed using either the rising or decaying phase solutions, $N_{\rm rise}$ and $N_{\rm decay}$, given by Equations (\ref{eq38}) and (\ref{eq47}), respectively, because the electron distribution is continuous at the peak of the sub-flare (see Equation~(\ref{eq46})).\\

\hoffset=-0.88truein
\begin{deluxetable}{ccccccccccc}
\tabletypesize{\scriptsize}
\tablecaption{Model Free Parameters \label{tbl-1}}
\tablewidth{0pt}
\tablehead{
\colhead{\textrm{sub-flare} \#}
& \colhead{$J_0$}
& \colhead{$\dfrac{E_*}{B_*}$}
& \colhead{$\hat S$}
& \colhead{$\hat C$}
& \colhead{$\mu$}
& \colhead{$\sigma$}
& \colhead{$\alpha$}
& \colhead{$\theta$}
& \colhead{$t_{\rm ad} \ ({\rm s})$}
& \colhead{$t_* \ (\textrm{MJD})$}
}
\startdata
$1$
&$7.94 \times 10^{38}$
&$0.085$
&$2.82 \times 10^{-20}$
&$0.2$
&$10^{5}$
&$3.43 \times 10^{9}$
&$6.15$
&$9.00$
&$1.75 \times 10^{5}$
&$55656.85$
\\
$2$
&$1.12 \times 10^{39}$
&$0.089$
&$1.47 \times 10^{-20}$
&$0.2$
&$10^{5}$
&$3.43 \times 10^{9}$
&$7.15$
&$4.65$
&$1.75 \times 10^{5}$
&$55660.85$
\\
\enddata
\end{deluxetable}

\hoffset=-0.88truein
\begin{deluxetable}{ccccccccc}
\tabletypesize{\scriptsize}
\tablecaption{Derived Parameters \label{tbl-2}}
\tablewidth{0pt}
\tablehead{
\colhead{\textrm{sub-flare} \#}
& \colhead{$\mathscr{N}_0$}
& \colhead{$B_{\rm pk} \ (\mu\textrm{G})$}
& \colhead{$\dfrac{E_{\rm pk}}{B_{\rm pk}}$}
& \colhead{$A_* \ ({\rm s}^{-1})$}
& \colhead{$w_{*}$}
& \colhead{$w_{\textrm{pk}}$}
& \colhead{$t_{\rm pk} \ ({\rm s})$}
& \colhead{$R_{\rm b} \ ({\rm cm})$}
}
\startdata
$1$
&$3.65 \times 10^{38}$
&$705.8$
&$1.84$
&$48.7$
&$58.9$
&$2.72$
&$7.08 \times 10^{5}$
&$1.75 \times 10^{15}$
\\
$2$
&$5.52 \times 10^{38}$
&$636.8$
&$3.18$
&$28.0$
&$56.1$
&$1.57$
&$5.48 \times 10^{5}$
&$1.75 \times 10^{15}$
\\
\enddata
\end{deluxetable}

\subsection{Light Curve}

In addition to the $\gamma$-ray spectra, we can also compute the theoretical light curve for the 2011 April super-flare by integrating our spectra above a photon energy of 100\,MeV. This is the same procedure carried out by \citet{buehler12} in their Figure~5, and therefore a comparison of our theoretical light curve with their observational data provides another interesting test of our time-dependent model. The light curve is obtained using a frequency (or energy) integration of the observed spectral flux, plotted in Figure~\ref{fig3}. The number flux, $F_{\rm tot}$, can be found by integrating Equation (\ref{eq52}) over frequency, which yields
\begin{equation}
F_{\rm tot}(t) = \frac{1}{4 \pi D^2}\int_{\nu_{\rm min}}^{\nu_{\rm max}}\frac{d\nu}{h\nu} \int_{x_{\rm min}}^{1/\sqrt{\hat{S}}}
N(x,t) P_\nu(\nu,x)dx
\ \propto \ {\rm \ s^{-1} \ cm^{-2}}
\ ,
\label{eq55}
\end{equation}
where $h$ is Planck's constant, $\nu$ is the photon frequency, $h\nu_{\rm min}=0.1\,$GeV and $h\nu_{\rm max}=100\,$GeV, and $D$ is the distance to the Crab nebula, which we set equal to 2\,kpc. We compute the theoretical light curve $F_{\rm tot}(t)$ for values of $t$ in the range between MJD~55662.5$-$55671.50. The observed flux outside this time range is near the quiescent level. When performing the integrations in Equation (\ref{eq55}), we set the electron distribution $N=N_{\rm rise}$ (Equation~(\ref{eq38})) for $t \le t_{\rm pk}$, and $N=N_{\rm decay}$ (Equation~(\ref{eq47})) for $t \ge t_{\rm pk}$. There are two separate blobs included in our model for the 2011 April super-flare, and therefore we must carry out the calculation in Equation~(\ref{eq55}) for each blob and then add the results together to obtain the final light curve.\\

\begin{figure}[h!]
\vspace{0.0cm}
\centering
\includegraphics[height=8cm]{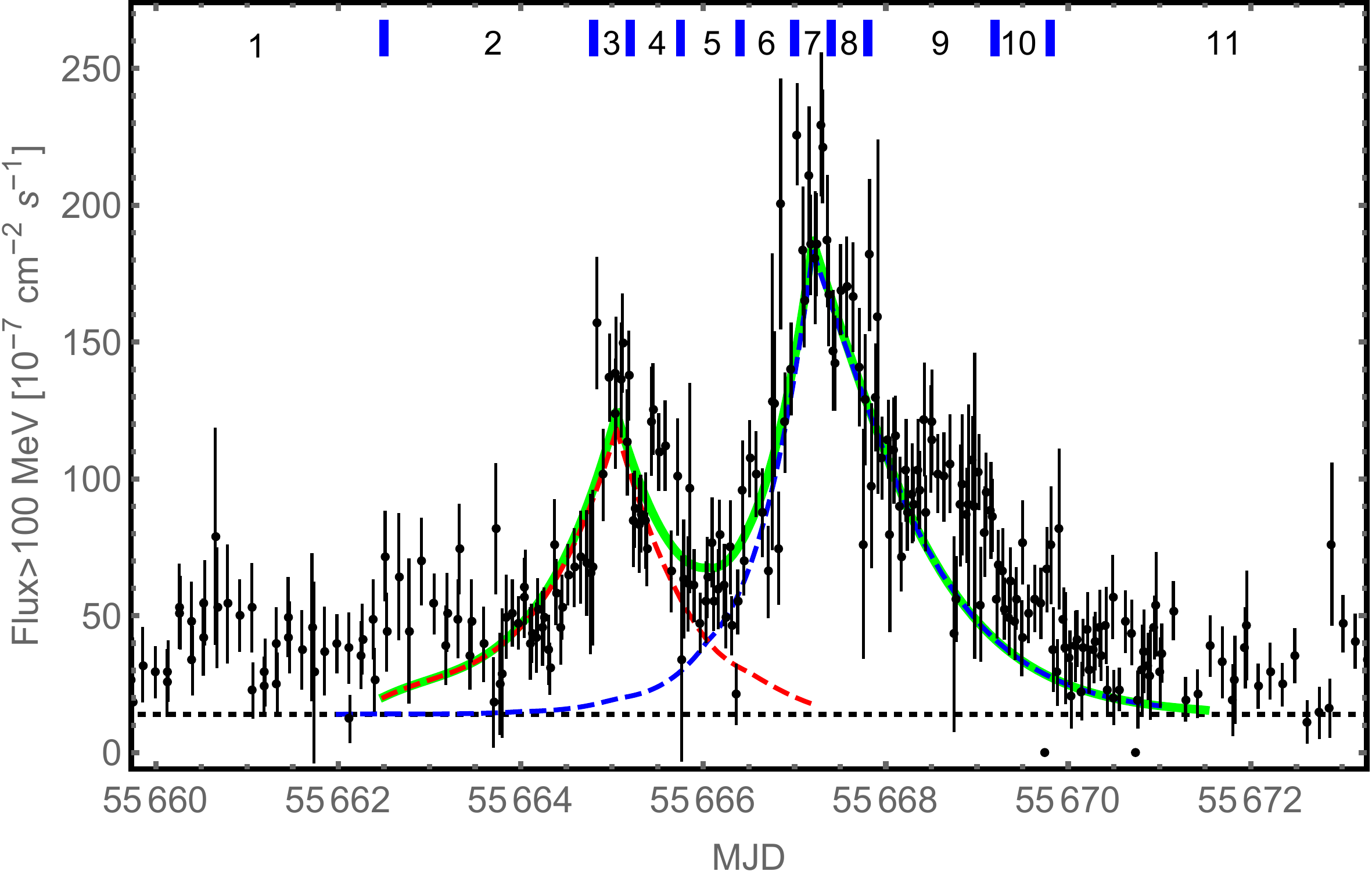}
\caption{Theoretical light curves computed using Equation (\ref{eq55}). The red and blue dashed curves represent the theoretical light curve for each individual sub-flare. The green solid curve represents the sum of each component. The black dashed line represents the 33 month average flux from the synchrotron nebula, obtained by integrating Equation (\ref{eq53}). The light curve data is taken from \citet{buehler12}.}
\label{fig6}
\end{figure}

In order to compare the light curve computed using our model (Equation~(\ref{eq55})) with the data presented by \citet{buehler12}, we must also include the contribution to the photon number flux due to the background nebular spectrum. This is accomplished by integrating the background nebula component (Equation (\ref{eq53})) with respect to frequency between $h\nu_{\rm min}=70\,$MeV and $h\nu_{\rm max}=100\,$GeV, which yields a background offset of $1.3 \times 10^{-6} \ {\rm s}^{-1} {\rm cm}^{-2}$. This amount is added to the light curves computed using Equation~(\ref{eq55}). In Figure \ref{fig6}, we compare our model with the observed light curve data. Individual light curves are plotted for each blob, corresponding to sub-flares 1 and 2, respectively, along with the total light curve for the entire flare event. We conclude that the model is able to reproduce the data.\\

\section{Discussion and Conclusion}

The Crab nebula $\gamma$-ray flares have presented considerable challenges to classical particle acceleration models. The classical synchrotron radiation-reaction (``burnoff'') limit and the sub-Larmor timescales in which electrons are inferred to have been accelerated place severe constraints on any model that attempts to reproduce the $\gamma$-ray flare spectra and explain the physical mechanisms responsible for these remarkable observations. The 2011 April super-flare was the brightest $\gamma$-ray transient observed from the Crab nebula, and it is therefore the focus of this study. We have presented a detailed derivation of a new time-dependent, analytical electron acceleration model that self-consistently reproduces the sequence of $\gamma$-ray spectra observed during the rising and decaying phases of each of the two sub-flare components detected by {\it Fermi}-LAT during the super-flare. The model also successfully reproduces the integrated $\gamma$-ray light curve, and provides new insight into the properties and mechanisms of the $\gamma$-ray transient.\\

One of the benefits of an analytic model such as the one developed here is that it gives one explicit control over the individual physical parameters that govern the processes of acceleration, losses, and escape. In addition, the model can be executed rapidly, which allows us to explore a wide variety of parameter values, profile functions, and initial conditions. Using our time-dependent model, we can confirm the suggestion made by \citet{kroon16} (based on a steady-state model) that second-order Fermi acceleration (momentum diffusion) is negligible during the flare. Hence the $\gamma$-ray emission is powered mainly by electrostatic acceleration in the strong electric fields produced via impulsive magnetic reconnection in the vicinity of the pulsar-wind termination shock.\\

Following the work of \citet{zrake16}, we have adopted a blob paradigm in which we interpret each sub-flare as emission from separate magnetically confined plasma structures from the cold pulsar wind which interact with the shock. The blobs are sufficiently separated spatially, but in such a manner in which each of their $\gamma$-ray fluxes are observable to an observer at Earth. The blob will not deteriorate before reaching the shock due to saturation of small amplitude instabilities \citep{zrake16}. Additionally, the duty cycle of the Crab nebula flares are likely the result of the frequency with which such blobs are produced \textit{and} the liklihood in which a blob produces $\gamma$-ray emission beamed towards Earth.\\

The analytic model presented here implements a ``profile function'' that quantifies the time-evolution of gains and losses. In this prescription, it is posited that the blob's interaction with the termination shock causes the electric and magnetic fields to evolve in time. Likewise, the shock-regulated escape mechanism also evolves in time since it is mediated by the magnetic field via the dependence on the particle's Larmor radius. We adopt an exponential rise/decay form for the profile function in agreement with prior studies \citep{buehler12}. The separate analytic solutions for the electron distribution function derived for the rising and decaying phases results in $\gamma$-ray synchrotron spectra that closely reproduce the observed spectral snapshots for the evolving flare. Furthermore, we also show that the light curve obtained by integrating the $\gamma$-ray spectra above $100\,$MeV reproduce the data plotted in Figure 5 of \citet{buehler12}. We note that there are suggestions of a possible third sub-flare (and a third plasma blob) contained in the light curve plotted by \citet{buehler12}. The model we have developed could be generalized to add such a third component, but we do not pursue that possibility here.\\ 

\vspace{-0.3cm}
\subsection{Energy Budget}

\begin{figure}[h!]
\captionsetup[subfigure]{labelformat=empty}
\subfloat[]{\includegraphics[scale=0.45]{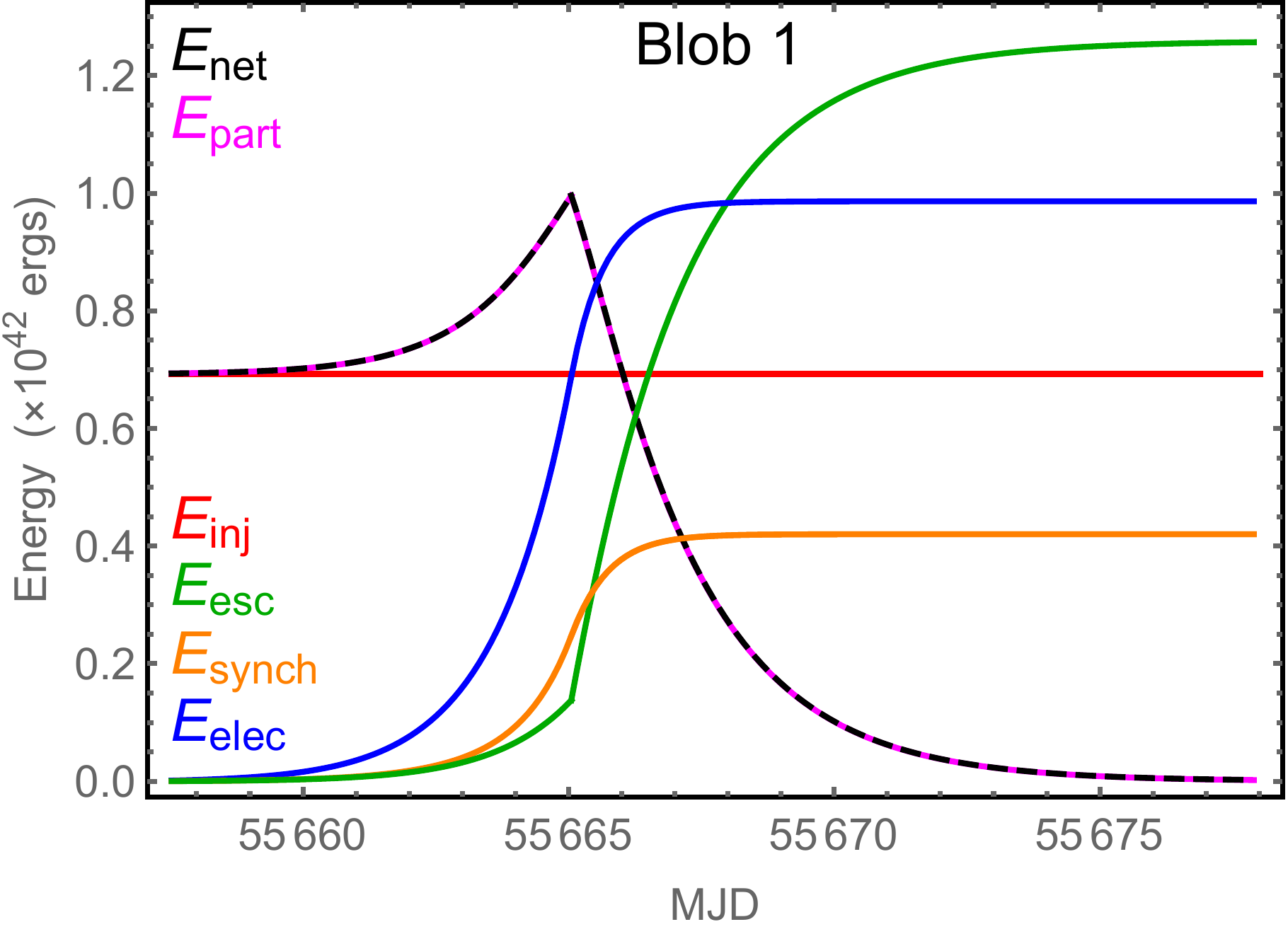}}
\subfloat[]{\includegraphics[scale=0.45]{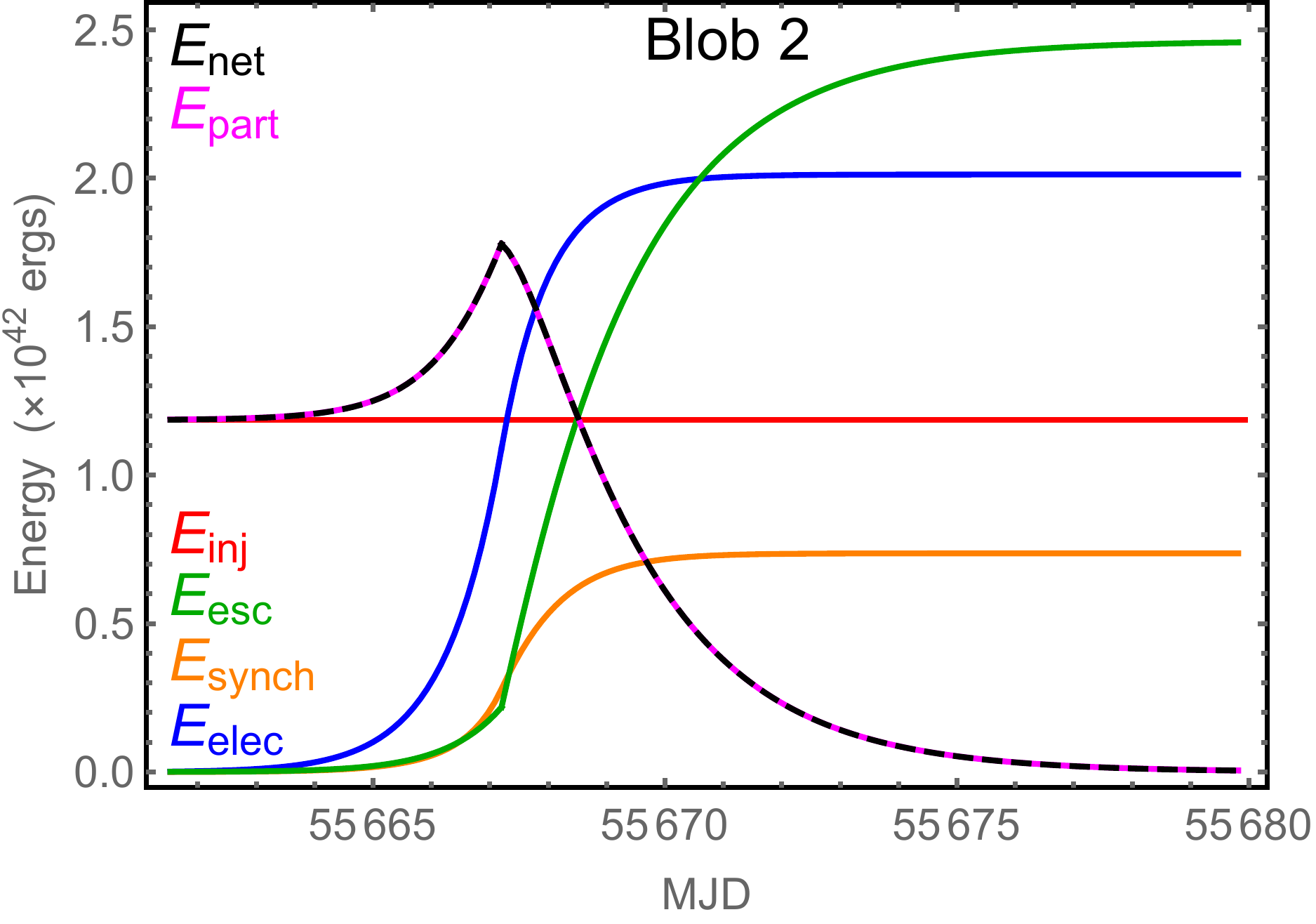}}
\caption{The cumulative energy channels are plotted as a function of the elapsed time $t$ for Blob 1 (left) and Blob 2 (right). The magenta curve (Equation (\ref{eq57})) is the total particle energy, $\mathscr{E}_{\rm part}(t)$, obtained by integrating the energy distribution, and the black dashed line denotes the net particle energy (Equation~(\ref{eq63})). The two curves agree as expected. We also plot the individual components due to escape (Equations (\ref{eq60}) and (\ref{eq61})), synchrotron losses (Equation~(\ref{eq59})), electrostatic acceleration (Equation~(\ref{eq58})), and the initial energy (Equation~(\ref{eq56})).}
\label{fig7b}
\end{figure}

An important test of the model is internal consistency and energy conservation. Below, we present the definitions of each energy channel and plot them in Figure \ref{fig7b}. We wish to quantify and compare the total energy of the evolving system. We can separate the total energy into particle energy, initial energy, electrostatic energy, synchrotron loss energy, and escaping energy due to shock-regulated escape. The total energy associated with the blob electrons at the beginning of the sub-flare is given by
\begin{equation}
\mathscr{E}_{\rm inj} = \int_0^{1/\sqrt{\hat{S}}} J_0 \, m_e c^2 \frac{\sqrt{x^2+1}}{\sigma \sqrt{2\pi}}e^{\frac{-(x-\mu)^2}{2\sigma^2}} dx
\label{eq56}
\ ,
\end{equation}
where $J_0$ is the normalization constant for the initial Gaussian distribution, and $m_e c^2 \sqrt{x^2+1}$ is the energy of an electron with dimensionless momentum $x$. The total energy of the electrons in the blob as a function of time, $t$, during the sub-flare is computed using
\begin{equation}
\mathscr{E}_{\rm part}(t) = \int_{x_{\rm min}(t)}^{1/\sqrt{\hat{S}}} m_e c^2 \sqrt{x^2+1}\ N(x,t)dx
\label{eq57}
\ ,
\end{equation}
where $x_{\rm min}$ is given by Equation (\ref{eq22}).\\ 

Next, we consider the energy gains and losses due to electrostatic acceleration, synchrotron losses, and particle escape. The cumulative energy pumped into the electrons in the blob via electrostatic acceleration between the beginning of the sub-flare at time $t_*$ and the current time $t$ is computed using the double integral
\begin{equation}
\mathscr{E}_{\rm elec}(t) = \int_{t_*}^{t} \int_{x_{\rm min}(t)}^{1/\sqrt{\hat{S}}} qcE(t') N(x,t') \, dx \, dt'
\label{eq58}
\ ,
\end{equation}
where $E(t')$ is the electric field as a function of time. The cumulative energy lost by the blob electrons due to synchrotron emission is given by
\begin{equation}
\mathscr{E}_{\rm synch}(t) = \int_{t_*}^{t} \int_{x_{\rm min}(t)}^{1/\sqrt{\hat{S}}} \frac{\sigmaT c}{6\pi} (x^2+1) B^2(t') N(x,t') \, dx \, dt'
\label{eq59}
\ ,
\end{equation}
where $\sigmaT$ is the Thomson scattering cross section and $B(t')$ is the time-dependent magnetic field.\\

During the rising phase of the sub-flare, the cumulative total energy lost from the blob due to shock-regulated particle escape is computed using
\begin{equation}
\mathscr{E}_{\rm SRE}(t) =  \int_{t_*}^{t} \int_{x_{\rm min}(t)}^{1/\sqrt{\hat{S}}} m_e c^2\sqrt{x^2+1} \, \frac{C(t)}{x} N(x,t') \, dx \, dt'
\label{eq60}
\ ,
\end{equation}
where $C(t)$ can be expressed as a time-dependent quantity given by the second expression in Equation (\ref{eq27}). Likewise, during the decaying phase of the sub-flare, we the total energy lost from the blob due to advective particle escape which is given by
\begin{equation}
\mathscr{E}_{\rm ad}(t) = \int_{t_*}^{t} \int_{x_{\rm min}(t)}^{1/\sqrt{\hat{S}}}m_e c^2\sqrt{x^2+1} \, \frac{ N(x,t')}{t_{\rm ad}} \, dx \, dt'
\label{eq61}
\ ,
\end{equation}
where $t_{\rm ad}$ is given by Equation (\ref{eq40}).\\

We can show that energy is conserved within the system during the rising phase by computing the expected net particle energy,
\begin{equation}
\mathscr{E}_{\rm net}^{\rm rise}(t) = \mathscr{E}_{\rm inj}+\mathscr{E}_{\rm elec}(t)
-\mathscr{E}_{\rm synch}(t)-\mathscr{E}_{\rm SRE}(t)
\label{eq62}
\ ,
\end{equation}
and comparing it with the total particle energy, $\mathscr{E}_{\rm part}(t)$, given by Equation (\ref{eq61}). Additionally, we can show that energy is conserved within the system during the decaying phase by computing
\begin{equation}
\mathscr{E}_{\rm net}^{\rm decay}(t) = \mathscr{E}_{\rm inj}+\mathscr{E}_{\rm elec}(t)
-\mathscr{E}_{\rm synch}(t)-\mathscr{E}_{\rm ad}(t)
\label{eq63}
\ ,
\end{equation}
and comparing it with $\mathscr{E}_{\rm part}(t)$. We carry out this comparison in Figure \ref{fig7b}, and show that energy is conserved as expected for each of the two blobs, corresponding to sub-flares 1 and 2.\\

Another interesting feature of the plots is that the cumulative energy radiated via synchrotron is about one third of the cumulative energy contained in the escaping electrons, which travel outward through the synchrotron nebula. We find that synchrotron accounts for about $24\%$ of the energy lost by the blob, which is consistent with observational estimates for the synchrotron efficiency \citep[e.g.,][]{abdo11}.  Hence the energy in the escaping electrons eventually contributes to the radio power generated farther out in the synchrotron nebula.\\

\subsection{Parameter Constraints}
\label{parameterconstraints}

Previous studies have demonstrated that electrostatic acceleration can provide an effective means of accelerating relativistic electrons to sufficiently high Lorentz factors in a region of reduced magnetic field \citep{cerutti14a,kroon16} to explain the production of $\gamma$-rays beyond the classical radiation-reaction (synchrotron burnoff) limit (Equation~(\ref{eq1})). This classical limit can be exceeded in non-ideal MHD conditions when the electric field exceeds the magnetic field, so that $E/B >1$. The quiescent state in the Crab nebula is most notably contrasted from the flaring state by an electric to magnetic field ratio of $E/B \ll 1$. This relation is consistent with our model parameters at the beginning of each $\gamma$-ray sub-flare, when we find that the initial electric and magnetic fields, $E_*$ and $B_*$, respectively, satisfy the relation $E_*/B_* \ll 1$ at the initial time $t=t_*$ (see Table~1). Conversely, at the peak of each sub-flare, we find that $E_{\rm pk}/B_{\rm pk} \sim\,$few, as reported in Table~2.\\ 

In our time-dependent, one-zone model, the SRE efficiency parameter, $w$, introduced in Equation~(\ref{eq27a}) describes the variation of the interaction between the blob and the shock, as well as variation of the effects of time dilation and magnetic obliquity. We report the initial value for the SRE efficiency parameter, $w_*$, in Table~2. Note that $w_* \gg 1$, implying that due to the highly relativistic upstream velocity, combined with the toroidal field geometry, shock-regulated escape is initially very ineffective, since the SRE escape timescale is large (see Equation~(\ref{eq27a})). This is consistent with the fact that the plasma blob has not yet encountered the shock. However, at the peak of the sub-flare, when the blob is in direct contact with the shock, we find that $w=w_{\rm pk} \sim 1$ (see Table~2), which indicates that the escape timescale is much smaller than at the start of the flare, for a given particle energy. The increase in the effectiveness of the SRE process near the peak of the flare reflects the reduction in the flow velocity as the blob crosses the shock, combined with a change in the magnetic obliquity. These effects tend to mitigate the effect of the particle acceleration by enhancing the escape of particles into the downstream region.\\

We find that as the blob passes through the termination shock, the dominant escape mechanism switches from shock-regulated escape on the upstream side to advective escape on the downstream side. The advective escape timescale, $t_{\rm ad}$, given by Equation (\ref{eq40}) depends on the downstream flow velocity, $v_{ds}$, and the blob size, $R_{\rm b}$. In our model, the value of $t_{\rm ad}$ is a free parameter which is reported in Table 1. Since $v_{ds}=c/3$, we can use Equation (\ref{eq40}) to compute the value of $R_{\rm b}$, which is reported in Table 2. We find that $R_{\rm b} \sim 10^{15}\,$cm. This result can be verified by examining the $\gamma$-ray light curve for the 2011 April super-flare. We observe that this event displayed a rise time of about one day, and we therefore expect the radius of the blob to be equal to about one light-day, and this is indeed the case. With the blob radius determined, we are in a position to re-examine the validity of the magnetic confinement assumption for the blob electrons. The highest energy particles (whose Larmor radii, $r_{\textrm{L}}$, might exceed the physical size of the blob) will be confined within the blob, provided the highly disordered magnetic field causes electrons to reflect off magnetic mirrors, or kinks. The mean distance between the mirrors, corresponding to the coherence length of the magnetic field, $\ell_{\textrm{coh}}$, is essentially equal to the blob radius, $R_{\textrm{b}}$. In this situation, the effective mean-free path for the electrons, $\ell$, is given by \citep{kroon16}
\begin{equation}
\dfrac{1}{\ell}=\dfrac{1}{r_{\textrm{L}}}+\dfrac{1}{\ell_{\textrm{coh}}}
\label{eq56a}
\ .
\end{equation}
This relation implies that $\ell < \ell_{\textrm{coh}} \sim R_{\textrm{b}}$ even when the electron Larmor radius exceeds the size of the blob, $r_{\textrm{L}} \gapprox R_{\textrm{b}}$.\\

It is also interesting to compute the magnetization parameter, $\sigma_{\textrm{pk}}$, at the peak of a given sub-flare, defined by
\begin{equation}
\sigma_{\textrm{pk}} = \dfrac{B^2_{\textrm{pk}}}{8\pi \mathscr{E}_{\textrm{pk}}}
\label{eq56b}
\ , 
\end{equation}
where $B_{\textrm{pk}}$ is the peak magnetic field strength and $\mathscr{E}_{\textrm{pk}} \equiv \mathscr{E}_{\textrm{part}}(t_{\textrm{pk}})$ is the total electron energy at the peak of the flare, computed using Equation (\ref{eq57}). For sub-flares one and two, we obtain $\sigma_{\textrm{pk}}=4 \times 10^{-4}$ and $\sigma_{\textrm{pk}}=2 \times 10^{-4}$, respectively. The small values for $\sigma_{\textrm{pk}}$ that we obtain imply that the plasma is weakly magnetized in the acceleration region. This is counterintuitive, since, as \citet{lyutikov16} point out, the particle acceleration powering the $\gamma$-ray flares from the Crab nebula is expected to occur in a region of strong magnetization, which is characteristic of explosive reconnection. It is difficult to completely resolve this paradox in the context of the one-zone model under consideration here. However, we argue that, at least qualitatively, the single zone treated here represents a spatial average over a reconnection cell, which includes a current sheet with low magnetic field strength at its core \citep[e.g.,][]{uzdensky11,cerutti13}, as well as a surrounding region of higher field strength. In this sense, the relatively low values for $\sigma_{\textrm{pk}}$ that we obtain correspond to the conditions near the current sheet. It is interesting to note that these results are consistent with those obtained by \citet{sironi13,sironi15}, who studied the acceleration of relativistic particles at relativistic collisionless shocks in weakly magnetized plasma in the context of gamma-ray bursts and pulsar-wind nebulae. Nonetheless, this is a question that certainly requires further analysis in future work.

\vspace{-0.3cm}
\subsection{Conclusion}
The $\gamma$-ray flares observed from the Crab nebula between 2007-2013 present serious challenges to classical particle acceleration mechanisms such as diffusive shock acceleration, which operates on MHD timescales and is therefore limited by synchrotron burnoff. However, this limit can be extended if one invokes strong electrostatic acceleration which is a plausible mechanism in the vicinity of pulsar-wind termination shocks, where rapid magnetic reconnection may occur due to the alternating magnetic polarity combined with the compression at the shock \citep{cerutti13}. In this paper we have used a transport equation to model the evolution of the electron energy distribution in a plasma blob that encounters the termination shock and experiences strong particle acceleration. The transport equation includes terms describing electrostatic acceleration, synchrotron losses, and shock-regulated escape. \citet{kroon16} found that electrostatic acceleration via shock-driven magnetic reconnection can provide sufficient energy to power the observed $\gamma$-ray flares.\\ 

In our model, we posit that each of the two sub-flares is due to emission from a magnetically-confined plasma blob that forms in the pulsar magnetosphere, upstream from the termination shock. The blobs are spatially separated and non-interacting, and their emission is powered by particle accelerating driven by magnetic reconnection occurring in the vicinity of the termination shock \citep{cerutti13}. Following \citet{buehler12}, we have therefore assumed that the physical variation of the magnetic and electric fields are correlated in time, following the same general evolutionary profile in time as the blob encounters the termination shock and passes through. This correlation is expressed by the ``profile function,'' $h(t)$, introduced in Equation (\ref{eq5}). The electrons in the blob initially experience a rising phase during which the particles tend to be recycled back across the shock by the shock-regulated escape mechanism. As discussed in Section \ref{TransEqn}, the qualitative behavior of the profile function can be deduced from observation of $\gamma$-ray light curves plotted in Figure \ref{fig6}. The profile function is an increasing exponential during the rising phase of the sub-flare, and a decreasing exponential during the decaying phase.\\

The short timescale variability observed during the $\gamma$-ray flares in our model is associated with the interaction of individual plasma blobs with the termination shock. However, in some alternative models, the short timescale variability is ascribed to the sweeping of a relativistic beam across the line of sight to the observer (Cerutti et al. 2012b). This scenario can work if the bulk Lorentz factor of the plasma producing the beamed emission is $\Gamma\sim 10$, which is implied by the solid angle $\Delta\theta = 0.03$\ sr as found by Cerutti et al. In the simplest version of a sweeping beam model, where the electron distribution does not change during the flare and the variability comes only from variations in the viewing angle, the implied spectral evolution would correspond to a simple diagonal translation of the logarithm of the observed flux relative to the logarithm of the observed photon frequency.  Further, both light curve and spectra should be completely symmetric if the beam is sweeping at a constant rate.  It is not clear if this type of spectral evolution is consistent with the observed light curve (Figure 7) and time-dependent gamma-ray spectra (Figure 4) of the Crab flares.  Observing these features (or not) could distinguish between a simple sweeping beam model and our model presented here.  A detailed analysis of these two models is, however, beyond the scope of this work.\\ 

Once the blob crosses over to the downstream side of the shock, the electrons experience a decaying phase in which the particle escape is dominated by energy-independent advection. We therefore introduce the assumption that the electric and magnetic fields follow the same general evolutionary profile in time as the blob encounters the termination shock and passes through, experiencing acceleration followed by a cooling phase in the downstream region, which is treated separately. The observed synchrotron emission in our model is predominantly generated in the region just downstream from the shock, because the radiative cooling timescale is longer than the acceleration timescale, except at the highest particle energies.\\ 
%

A novel feature of the model developed here is the two-phase particle escape mechanism, in which two different mechanisms dominate the particle escape before and after the peak of the $\gamma$-ray emission. When the blob is not yet fully through the termination shock plane, particles egress from the blob primarily via shock-regulated escape. However, after the peak of the sub-flare, when the blob has passed through the shock plane, the particle escape is dominated by advective escape, which is energy-independent. We find that the advective escape timescale, $t_{\rm ad}$, is consistent with the downstream diffusion velocity, $v_{ds}$, of the blob in the post-shock region (see Equation (\ref{eq40})). Once particles escape from the plasma blob, the strength of the magnetic field they experience in the ambient nebula is $\sim 200\,\mu$G, which is less than half the value during the peak of the flare. Computations of the afterglow spectra (not presented here) demonstrate that the emission produced by the electrons that have escaped from the blob does not contribute significantly to either the spectrum or the light curve for the $\gamma$-ray flares considered here.\\

For a given magnetic field strength, assuming a one-zone flare model in which the $\gamma$-rays are produced by synchrotron by a blob interacting with the termination shock, we have obtained unique solutions for the electron distributions at every point in time during the two sub-flares that comprise the 2011 April super-flare observed from the Crab nebula. These solutions are plotted in Figures \ref{fig4} and \ref{fig5}. It is important to emphasize that even though our model includes some simplifications and idealizations, the resulting electron energy distributions are necessarily correct, whether or not one wishes to challenge any aspects of our model. Hence, the work presented here really has two components. The first is the determination of the electron distributions associated with the 2011 April super-flare, and the second is the development of an approximate analytical transport model that successfully explains the production of those electron distributions.\\ 

Our detailed results confirm that shock-driven magnetic reconnection continues to be the most promising mechanism for accelerating electrons at the termination shock to very high energies in sub-Larmor timescales \citep{cerutti13}. Correspondingly, we find that electrostatic acceleration provides the vast majority of the observed power generated during a sub-flare and that a flare is characterized by non-ideal MHD conditions in which the $E/B$ ratio can be as high as $\sim 3$ during peak emission. Interestingly, the less luminous sub-flare was found to have $E/B=1.8$, almost a factor of two less than the brighter sub-flare, as well as a smaller population of electrons contributing to the emission. However, it should be noted that there remains some uncertainty about whether electrostatic acceleration alone can provide a complete explanation for the observations \citep{coroniti90,olmi15}.\\

We gratefully acknowledge useful several comments from the anonymous referee which helped us to significantly improve the manuscript.

J.J.K. was supported at NRL by NASA under contract S-15633Y.  J.D.F. was supported by the Chief of Naval Research.\\

\appendix
\section{Appendix}

We can show that a given photon spectrum can be produced via synchrotron emission from a family of different electron energy distribution that are correlated with the magnetic field strength. In other words, the same spectrum can be produced by electron populations residing in two different magnetic field strengths if the electron energy distributions are appropriately related to each other. We start with the synchrotron convolution integral in which a photon spectrum is computed from an arbitrary particle distribution function $N(\gamma)$ and magnetic field $B$ which is given by \citep{rybicki79}
\begin{equation}
L_{\nu}(\nu)=\int_{1}^{\infty}\frac{B q^3\sqrt{3}}{m_e c^2} R\left(\frac{4\pi m_e c \, \nu}{3\gamma^2 q B}\right)N(\gamma)d\gamma \ \ \propto {\rm erg} \ {\rm s}^{-1} \ {\rm Hz}^{-1}
\label{eq64}
\ ,
\end{equation}
where the function $R_{\rm b}$ is given by Equation (\ref{eq50}). Next we ask what electron distribution, $N'(\gamma')$, would be required in order to emit the same synchrotron spectrum, but in a different magnetic field, $B'$. This would imply that
\begin{equation}
\frac{q^3 B\sqrt{3}}{m_e c^2} R\left(\frac{4\pi m_e c \nu}{3\gamma^2 q B}\right)N(\gamma)d\gamma=\frac{q^3 B'\sqrt{3}}{m_e c^2} R\left(\frac{4\pi m_e c \nu}{3\gamma'^2 q B'}\right)N'(\gamma')d\gamma'
\label{eq65}
\ ,
\end{equation}
from which we conclude that
\begin{equation}
\gamma^2 B=\gamma'^2 B' \qquad \Rightarrow \qquad \gamma'=\gamma \sqrt{B \over B'}
\label{eq66}
\ .
\end{equation}
This in turn implies that the particle distribution functions are related to each other via
\begin{equation}
N'(\gamma')=\sqrt{\frac{B'}{B}}N\left(\gamma \sqrt{B' \over B}\right)
\label{eq67}
\ .
\end{equation}
This relationship has significant implications related to the conclusions we present in the paper. We have derived an exact solution for the time-dependent electron distribution assuming that the magnetic field strength is known. In fact, in the case of the Crab nebula, the field is fairly well constrained to lie in the range $100 \, \mu {\rm G} \lapprox B \lapprox 500 \, \mu {\rm G}$ \citep{aharonian04}. If the magnetic field strength is varied from the dependence we have assumed here (Equation (\ref{eq7})), then Equation (\ref{eq66}) implies that the electron distribution will transform homologously as a function of the Lorentz factor transformation. However, the shape of the electron distribution, will not change, and the determination of that shape is one of the major findings of this paper.\\


\begin{thebibliography}{}

\bibitem[Abdo et al.(2011)]{abdo11} Abdo, A., et al.\ 2011, Science, 331, 739

\bibitem[Achterberg et al.(2001)]{achterberg01} Achterberg, A., et al.\ 2001, MNRAS, 328, 393

\bibitem[Aharonian et al.(2004)]{aharonian04} Aharonian, F., et al.
2004, \apj, 614, 897

\bibitem[Becker(1992)]{becker92} Becker, P.~A.\ 1992, \apj, 397, 88

\bibitem[B\"uhler et al.(2012)]{buehler12} B\"uhler, R., et al.\ 2012, \apj, 749, 26

\bibitem[B{\"u}hler \& Blandford(2014)]{buehler14}
B{\"u}hler, R., \& Blandford, R. 2014, Reports on Progress in
Physics, 77, 066901

\bibitem[Cerutti et al.(2012a)]{cerutti12a} Cerutti, B., Uzdensky,
D.~A., \& Begelman, M.~C. 2012a, \apj, 746, 148

\bibitem[Cerutti et al.(2012b)]{cerutti12b} Cerutti, B., Werner,
G.~R., Uzdensky, D.~A., \& Begelman, M.~C. 2012b, \apjl, 754, L33

\bibitem[Cerutti et al.(2013)]{cerutti13} Cerutti, B., Werner, G.~R.,
Uzdensky, D.~A., \& Begelman, M.~C.\ 2013, \apj, 770, 147

\bibitem[Cerutti et al.(2014a)]{cerutti14a} Cerutti, B., Werner, G.~R., Uzdensky, D.~A., \& Begelman, M.~C.\ 2014a, ApJ, 782, 104

\bibitem[Cerutti et al.(2014b)]{cerutti14b} Cerutti, B., Werner, G.~R., Uzdensky, D.~A., \& Begelman, M.~C.\ 2014b, PhPl, 21, 6501

\bibitem[Coroniti(1990)]{coroniti90} Coroniti, F.~V.\ 1990, ApJ, 349, 538

\bibitem[Crusius \& Schlickeiser(1986)]{crusius86} Crusius, A., \& Schlickeiser, R.\ 1986, A\&A, 164, L16

\bibitem[Ellison et al.(1990)]{ellison90} Ellison, D.~C., Jones, F.~C., \& Reynolds, S.~P.\ 1990, ApJ, 360, 702

\bibitem[Gaensler \& Slane(2006)]{gaensler06} Gaensler, B.~M., \& Slane, P.~O.\ 2006, ARA\&A, 44, 17

\bibitem[Gallant et al.(1992)]{gallant92} Gallant, Y.~A., et al.\ 1992, ApJ, 391, 73

\bibitem[Hester(2008)]{hester08} Hester, J.~J. 2008, ARA\&A, 46, 127

\bibitem[Kennel \& Coroniti(1984)]{kennel84} Kennel, C. F., \& Coroniti, F.~V.\ 1984, ApJ, 283, 710

\bibitem[Komissarov(2013)]{komissarov13} Komissarov, S.~S.\ 2013, MNRAS, 428, 2459

\bibitem[Kroon et al.(2016)]{kroon16} Kroon, J.~J., Becker, P.~A.,
Finke, J.~D., \& Dermer, C.~D.\ 2016, \apj, 833, 157

\bibitem[Lemoine \& Waxman(2009)]{lemoine09} Lemoine, M., \& Waxman, E.\ 2009, JCAP, 2009, 9

\bibitem[Lyubarsky(2003)]{lyubarsky03} Lyubarsky, Y.~E.\ 2003, MNRAS, 345, 153

\bibitem[Lyutikov et al.(2016)]{lyutikov16} Lyutikov, M., Komissarov, S. S., \& Porth, O. \ 2016, \mnras, 456, 286

\bibitem[Montani \& Bernardini(2014)]{montani14} Montani, G., \& Bernardini, M.~G. \ 2014, Physics Letters B, 739, 433

\bibitem[Olmi et al.(2015)]{olmi15} Olmi, B., Del Zanna, L.,
Amato, E., \& Bucciantini, N.\ 2015, MNRAS, 449, 3149

\bibitem[Rees \& Gunn(1974)]{rees74} Rees, M.~J., \& Gunn, J.~E.\ 1974, MNRAS, 167, 1

\bibitem[Rybicki \& Lightman(1979)]{rybicki79} Rybicki, G. B., \& Lightman, A. P. 1979, Radiative Processes in Astrophysics (New York: Wiley)

\bibitem[Sironi et al.(2013)]{sironi13} Sironi, L., Spitkovsky, A., \& Arons, J. \ 2013, ApJ, 771, 54

\bibitem[Sironi et al.(2015)]{sironi15} Sironi, L., Keshet, U., \& Lemoine, M. \ 2015, SSRv, 191, 519

\bibitem[Striani et al.(2011)]{striani11} Striani, E., Tavani, M., \& Piano, G., et al.\ 2011, ApJL, 741, L5

\bibitem[Uzdensky et al.(2011)]{uzdensky11} Uzdensky, D.~A., Cerutti, B., \& Begelman, M. C., 2011, ApJ, 737, L40

\bibitem[Uzdensky \& Spitkovsky(2014)]{uzdensky14} Uzdensky, D.~A., \& Spitkovsky, A.\ 2014, ApJ, 780, 3

\bibitem[Zrake(2016)]{zrake16} Zrake, J.,\ 2016, ApJ, 823, 39

\end{thebibliography}
\end{document}